\documentclass[a4paper,12pt]{article}

\pdfoutput=1

\makeatletter
\def\ps@pprintTitle{%
 \let\@oddhead\@empty
 \let\@evenhead\@empty
 \def\@oddfoot{}%
 \let\@evenfoot\@oddfoot}
\makeatother

\newdimen\plotwidthSDM \newdimen\plotwidthHMOneD \newdimen\plotwidthHMTwoD
\DeclareTextCommand{\_}{OT1}{\leavevmode \kern.06em\vbox{\hrule width.6em}}

\usepackage{tikz}
\usetikzlibrary{shapes,arrows,positioning,matrix}

\usepackage{a4wide}
\usepackage{amssymb}
\usepackage{amsmath}
\usepackage{amsfonts}
\usepackage{graphicx}
\usepackage{caption}
\usepackage{booktabs}
\usepackage{subcaption}

\usepackage{listings}
\usepackage{color}
\usepackage{hyperref}
\usepackage{xspace}
\usepackage{MnSymbol} 

\definecolor{gray}{gray}{0.6}
\definecolor{darkgray}{gray}{0.4}
\definecolor{lightblue}{rgb}{0.92,0.92,1}

\lstdefinelanguage{pseudocode}{
    keywords={While,If,For,each,Else,from,to,for,in,End,Add,add,Replace,Apply,not,and,or,is,Redo,Next},
    keywordstyle={\color{black}\bfseries},
    mathescape=true
}
\lstdefinelanguage{config}{
    keywords=[1]{setup,parameter_space,algorithm},
    keywords=[2]{
        version,mode,point_processor,
        unit_length,chain_length,concurrent_processors, 
        workers,authkey,template,
        files,out_columns,
        par_names,
        var_names,
        data_names,file_columns,
        bound_count,
        likelihood,min_likelihood,likelihood_steps,
        loading_filter,projection_selector,
        projection_count,extrapolated_projections,
        symmetry_count,@include,point_count
    },
    sensitive=false,
    comment=[l]{\#},
    morecomment=[l][\color{black}]{=}, 
    morecomment=[s][\color{black}]{\%(}{s}, 
    string=[b]",
    keywordstyle=[1]{\color{gray}\bfseries},
    keywordstyle=[2]{\color{black}\bfseries},
    emphstyle=[2]{\color{gray}\bfseries},
    emphstyle=[1]{\color{black}\bfseries},
    commentstyle={\color{darkgray}\bfseries}
}

\newcommand{\R}{\textsuperscript{\textregistered}}
\newcommand{\cliarg}[1]{\texttt{\small{-}-#1}}
\newcommand{\fncliarg}[1]{\texttt{\scriptsize{-}-#1}}
\newcommand{\code}[1]{\texttt{\small #1}}
\newcommand{\fncode}[1]{\texttt{\scriptsize #1}}
\newcommand{\opt}[1]{\texttt{\bfseries\small #1}}
\newcommand{\optsec}[1]{\texttt{\bfseries\small [{\color{gray}#1}]}}

\newcommand{\program}{T3PS\xspace}
\newcommand{\programfilename}{t3ps\xspace}
\newcommand{\programlong}{Tool for Parallel Processing in Parameter Scans\xspace}
\newcommand{\programversion}{1.0\xspace}

\newcommand{\programpackagename}{T3PS-1.0.tar.gz}
\newcommand{\programpackagefolder}{T3PS-1.0}

\begin{document}

\begin{titlepage}

\begin{center}
{
\bf\LARGE
\program v\programversion: \programlong
}
\\[8mm]
Vinzenz Maurer
\footnote{E-mail: \texttt{vinzenz.maurer@unibas.ch}}
\\[1mm]
\end{center}
\vspace*{0.50cm}
\centerline{\it Department of Physics, University of Basel,}
\centerline{\it Klingelbergstr.~82, CH-4056 Basel, Switzerland}
\vspace*{1.20cm}

\begin{abstract}
\noindent
\program is a program that can be used to quickly design and perform parameter scans while easily taking advantage 
of the multi-core architecture of current processors. It takes an easy to read and write parameter scan 
definition file format as input. Based on the parameter ranges and other options contained therein, it distributes
the calculation of the parameter space over multiple processes and possibly computers. The derived data is saved 
in a plain text file format readable by most plotting software. The supported scanning strategies include: grid 
scan, random scan, Markov Chain Monte Carlo, numerical optimization. Several example parameter scans are shown and 
compared with results in the literature.

\end{abstract}

\end{titlepage}

\setcounter{footnote}{0}

\newpage
\tableofcontents
\newpage

\noindent \textbf{Disclaimer:}

\noindent \program is published under the GNU General Public License
\footnote{http://www.gnu.org/licenses/lgpl.html}. This means that you can use it for free.
We have tested this software and its results, but we can't guarantee that this software works correctly or that 
the results derived using it are correct.

\noindent
If you encounter any problem or have a question, feel free to contact the author by e-mail.

\newpage

\section{Introduction}

\program is a Python program that facilitates swift and versatile implementation of parameter scans on 
multiple CPUs and/or computers. 

Its main focus lies in delegation of tasks (``\textit{points}'' in parameter space) to sub-processes 
(``\textit{\textit{point processors}}'', see \ref{sec:pointprocessor}) that calculate information about these
\textit{points}. These sub-processes can run in parallel on one or more computers. To this end, each 
evaluation of a point works the following way:

Each \textit{point} is characterized by a list of values with fixed length. These values are substituted into a 
\textit{template file} in pre-determined positions. The file is then saved into a separate temporary folder and
its path is given to a \textit{point processor}, which returns the results derived from the 
\textit{substituted template file} as another list of numbers. These numbers are then saved together with the 
previous list separated by tabulators (``\textbackslash t'') to a result file as a single line. Calculations can 
be interrupted and resumed at any time\footnote{\program will try to resume at the exact step it was interrupted 
at, but this may not always be possible due to constraints of the algorithm or implementation.}.

For determining which points should be processed, the following strategies are implemented in \program:
\begin{itemize}
    \item Simple scan: the user directly specifies a grid, basic probability distributions or an explicit list 
        of points that shall all be processed.
    \item Markov Chain Monte Carlo (MCMC) algorithm: the user specifies ranges or discrete sets of parameter
        values, a propagation likelihood\footnote{In the case of uniformly distributed parameters, this is 
        exactly the target probability distribution function.} and the number of points to calculate to obtain
        a sample for the likelihood in question.
    \item Optimizing scan: the user specifies ranges or discrete sets for the values of the defined parameters.
        Using an evolutionary algorithm, the global maximum of a user-supplied fitness function is searched for.
    \item Exploring scan: the user specifies a grid of points and a likelihood function. The points will be 
        calculated gradually while aiming to find more likely or more interesting points earlier than others.
\end{itemize}
Additionally, \program supports the following auxiliary modes:
\begin{itemize}
    \item Test mode: point values are provided by hand by the user. This is useful for checking the implementation 
        of a given \textit{point processor} and scan definition before committing to a lengthy parameter scan.
    \item Worker mode: This mode is used only in setups involving multiple computers. In this mode, the 
        running instance of \program does not come up with new points on its own but rather waits for a
        ``manager'' process to supply them.
\end{itemize}

This manual is organized the following way: after a quick introduction to \program in sec.~\ref{sec:quickstart}, 
the general structure, strategies and concepts of typical parameter scans are shown in \ref{sec:overview}. 
In sec.~\ref{sec:scandefinition}, the scan definition format is introduced and all possible definition directives 
are listed and explained. The command line interface is detailed in sec.~\ref{sec:CLIarguments}.
In sec.~\ref{sec:gridcomputing}, the implementation of multi-computer calculations is described.
Finally in sec.~\ref{sec:providedprocessors}, the already implemented and included \textit{point processors} are 
documented. Everything is put together in several example scans in sec.~\ref{sec:examples}.

\subsection{System Requirements}

The following software packages must be installed for \program to run:
\begin{itemize}
    \item Python 2.7 or compatible, e.g.\ PyPy 1.5 or higher\footnote{Note that two of the examples in 
    sec.~\ref{sec:Example_ChargedLeptons} use the Python module ``SciPy'', which is as of writing this manual 
    not fully available for PyPy. It may thus not run completely unchanged and out of the box compared to 
    using standard CPython.}
    \item Unix-like operating system
\end{itemize}

\subsection{Installation and Quick Start}\label{sec:quickstart}

\program is usable right after extracting its archive file and possibly making the main file executable: 
\begin{lstlisting}[language=sh,escapechar=|,keywords={},emph={$}]
$ tar xvfz |\programpackagename|
$ cd |\programpackagefolder|
$ chmod u+x src/|\programfilename|
$ ./src/|\programfilename|scan_definition.scan
\end{lstlisting}
For easier usage, it can also be properly installed using 
\begin{lstlisting}[language=sh,escapechar=|,keywords={},emph={$}]
$ make install
\end{lstlisting}
which (after a prompt to the user) will copy \program to an installation folder and will create a link such that 
\program can be launched by simply typing 
\begin{lstlisting}[language=sh,escapechar=|,keywords={},emph={$}] 
$ |\programfilename|
\end{lstlisting}

As an introduction, we will now show how to perform a very basic scan. We begin with the scan definition file 
called ``\textbf{QuickStart.scan}'': 
\begin{lstlisting}[language=config,moreemph={par_x,par_y,program},morekeywords={[1]SimpleProcessor}]
[setup]
point_processor = processors/SimpleProcessor.py
template = QuickStart.function

[SimpleProcessor]
# do a scan with the `basic calculator' 
#  with basic mathematical functions
program = bc --mathlib

[parameter_space]
# two parameters called x and y
par_names = x, y
# define each parameter to take 100 different 
#  values going from -1 to 1
par_x = interval(-1, 1) with count = 100
par_y = interval(-1, 1) with count = 100
\end{lstlisting}
The \textit{point processor} called \code{SimpleProcessor} (included in the \program package) will use the basic 
calculator \code{bc} to calculate the mathematical function \mbox{$\sin(x^2 + y) \cos(y^2 + 3 x)$} by inserting the 
$x$ and $y$ values into the \textit{template file} with the name ``\textbf{QuickStart.function}'' and the
following content\footnote{In \code{\footnotesize bc}, \code{\footnotesize s} and \code{\footnotesize c} 
denote the $\sin$ and $\cos$ functions respectively.}:
\begin{lstlisting}[showlines=true]
s( ($x)^2 + ($y) ) * c( ($y)^2 + 3 * ($x) )

\end{lstlisting}
The only thing left to do for the scan is launching \program using 
\begin{lstlisting}[language=sh,escapechar=|,keywords={},emph={$}]
$ |\programfilename|QuickStart.scan
\end{lstlisting}
After a short while in the command line interface of \program, the resulting data set can be plotted in 
Wolfram Mathematica\R \cite{Mathematica_7.0} using the code 
\begin{lstlisting}[language=Gnuplot,keywords={ListContourPlot,Import},emph=points]
points = Import["QuickStart.scan.data", "Table"];
ListContourPlot[points[[All, {1,2,3}]]]
\end{lstlisting}
and in gnuplot \cite{Gnuplot_4.4} using the code 
\begin{lstlisting}[language=Gnuplot]
splot "QuickStart.scan.data" using 1:2:3
\end{lstlisting}

For examples that go more in-depth and show more features, see sec.~\ref{sec:examples}.

\section{Overview}\label{sec:overview}

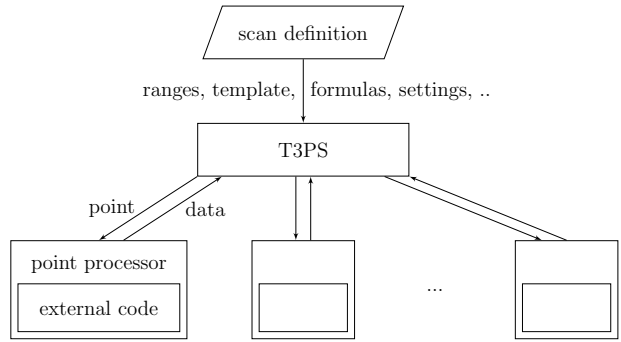
\begin{figure}
\begin{center}
\scalebox{0.65}{
    \begin{tikzpicture}[node distance=1.3cm,>=latex',line width=0.5pt,
        block/.style = {draw, shape=rectangle, align=center},
        tblock/.style = {draw, shape=trapezium, trapezium left angle=70, 
                 trapezium right angle=110, align=center},
    ]
    \node [tblock] (scandef) {\parbox[c][0.8cm]{3cm}{\centering scan definition}};
    \node [block, below=of scandef] (program) {\parbox[c][0.8cm]{4cm}{\centering \program}};
    \node [block, below=of program] (process2) {
        \parbox[c][0.7cm]{3cm}{\phantom{\tiny point processor}}\\
        \tikz\node[draw,shape=rectangle]{\parbox[c][0.7cm]{3cm}{\phantom{\tiny external code}}};
    };
    \node [block, below=of program, left=of process2] (process1) {
        \parbox[c][0.7cm]{3cm}{\centering point processor}\\
        \tikz\node[draw,shape=rectangle]{\parbox[c][0.7cm]{3cm}{\centering external code}};
    };
    \node [below=of program, right=of process2] (process3) {...};
    \node [block, below=of program, right=of process3] (process4) {
        \parbox[c][0.7cm]{3cm}{\phantom{\tiny point processor}}\\
        \tikz\node[draw,shape=rectangle]{\parbox[c][0.7cm]{3cm}{\phantom{\tiny external code}}};
    };
    \draw [->] (scandef) to node[left] {ranges, template, } node[right] {formulas, settings, ..} (program);
    
    \draw [->] (program.south west) to node[left] {point\;} (process1.north);
    \draw [transform canvas={xshift=5mm},->] (process1.north) to node[right] {\;data} (program.south west);
    \draw [transform canvas={xshift=-1.5mm},->] (program) to (process2);
    \draw [transform canvas={xshift=1.5mm},->] (process2) to (program);
    \draw [transform canvas={xshift=-5mm},->] (program.south east) to (process4.north);
    \draw [->] (process4.north) to (program.south east);
    \end{tikzpicture}
}
\end{center}
    \caption{
        Flow chart representing the overall algorithmic structure of a scan in \program. The main program 
        takes the information provided by the scan definition file and distributes the work to a certain number of 
        concurrently running \textit{point processors}, one point each. These in turn calculate data using some 
        (possibly external) code and return the relevant data.
    }
\end{figure}

\subsection{Scanning Strategies}

We will now go into more detail how each scanning strategy behaves and is (roughly) implemented. For more detailed 
sketches of the actual implementations, see app.~\ref{app:algorithms}.

\paragraph{Scan}
This mode simply processes a defined set of points in the specified parameter space. There are three basic 
modes for specifying it.

In the first mode, the user specifies the parameter space as a grid defined as the direct product of finite ranges 
for each parameter. It is then scanned over by simple iteration over all points on the grid. This mode is 
called ``grid'' mode.

If one or more parameter range is continuous, but the full space is still a direct product, the second 
mode applies. Here a user-specified number of points in parameter space is randomly chosen (according to their
distribution) and calculated. \program calls this mode ``scatter''.

In the third mode, instead of a parameter space as direct product of ranges, the user can supply a file in 
tabulator separated values (TSV) format that specifies the set of possible parameter value combinations that
constitutes the parameter space. This mode is called ``file'' mode.

\paragraph{MCMC}
This mode uses a standard Metropolis-Hastings algorithm \cite{MetropolisEtAl53,Hastings70} to draw
random samples of points from a more complicated probability distribution. This is also known as a Markov 
Chain Monte Carlo (MCMC) method.

The range for each parameter can either be given a set of discrete and equally probable values or as 
continuous values following some (simple) probability distribution. In both cases, the user can supply a step 
size for a Gaussian proposal density around the parameter value of the previous iteration.
If no step size is given, the specific point coordinate is chosen at random from their prior probability 
distribution given by the parameter range definition.

Once a Markov chain reaches a user-supplied number of sample points (not taking into account non-trivial stay
counts), it terminates.

For a sketch in pseudocode of the implementation in \program, see sec.~\ref{sec:mcmc_algorithm}.

\paragraph{Optimize}
This mode implements\footnote{This is also implemented in the NMinimize/NMaximize functions of Wolfram 
Mathematica\R.} a differential evolution algorithm \cite{StornPrice97} to optimize a user-supplied fitness 
function. Here, \program tries to find the global fitness maximum by (starting from a randomly found or 
user-supplied initial population of points) repeatedly applying an evolutionary approach that yields a new 
-- better or same quality -- population in each iteration.

It stops once the best found fitness has not changed by more than a user-supplied threshold value for a 
user-supplied number of iterations. If the full parameter space has only a finite amount of points, this mode 
caches calculation results and thus never calculates points twice.

For a sketch in pseudocode of the implementation in \program, see sec.~\ref{sec:optimize_algorithm}.

\paragraph{Explorer}
This mode implements a custom swarm type algorithm acting on a discrete finite grid, analogous to the plain 
``scan'' mode.
Starting from randomly found or user-supplied points and emanating from further calculated points, the neighbor 
points (relative to the grid) of a subset of points are calculated in each iteration, where points with a higher 
value of a user supplied ``likelihood'' function take precedence over ones with lower values. \program keeps track 
of all calculated points and no point is processed twice (some memory usage optimizations can change this slightly).
Since on a finite grid calculation time is also finite, this is just a re-ordering compared to doing the same
calculation in ``scan'' mode. However, if one is only interested in points satisfying a particular needed level
of likelihood, this method can save time when dealing with extensive parameter spaces (depending on the starting
points), while still benefiting from the regularity and controllable density of grid-form parameter spaces.

An additional feature is the specification of projections over which a function shall be maximized (not 
necessarily the ``likelihood'' function).
When the user specifies such a projection, \program groups up all known points by two (user supplied) coordinates 
$x$, $y$ and third coordinate $z$ (default is the ``likelihood''). Then the points with maximal $z$ coordinate 
$z_{\max}$ for each point in the $x$, $y$ plain are used first as site for the calculation of further neighbor 
points. In addition to simple neighbors, these optimal points are interpolated and extrapolated to smoothen out the 
projected $x$-$y$-$z_{\max}$ graph\footnote{Thus projections work best if the $z$ coordinate function is 
sufficiently linear over the scale of the grid spacing.}.

More specifically, the algorithm has three states. In \textit{state one}, only one point with maximal $z$ per $x$ 
and $y$ is found and has his neighbor points and extra-/interpolated pseudo-neighbors calculated. If the 
algorithm does not find any new points this way, it goes into \textit{state two} and instead looks for the 
points maximizing $z$ while being on the boundary of the point set, i.e.\ ones that have some missing neighbors, 
and calculates those missing neighbors. In the case, where the algorithm does not find any new points in 
\textit{state two}, it goes into \textit{state three}, where projections are not considered and just a set of 
pre-defined number of boundary points with sufficient likelihood has their neighbors calculated.

If there are no projections defined, the algorithm always automatically goes into \textit{state three}.
If there are projections defined, every time new points are found the algorithm reverts back to \textit{state one}.

Moreover, in the case where approximate symmetries of the parameters are known under which the results are
almost invariant, these can also be specified and used in \textit{state one} to further refine the 
$x$-$y$-$z_{\max}$ graphs by applying the corresponding transformations to the points with maximal $z$ coordinate
in each plain.

Note that there is no automatic convergence analysis and users must decide for themselves whether or not the 
calculation has run long enough. \program only calculates some possibly helpful indicators involving the change in
the projection plains and other characteristics between each iteration.

For a sketch in pseudocode of the implementation in \program, see sec.~\ref{sec:explorer_algorithm}.

\paragraph{Worker}
As a worker instance, \program will not start any calculations on its own, but will listen on a user-supplied 
network port (default: 31415) for calculation requests from \program manager instances. To make sure only
appropriate processes make these requests a shared secret also referred to as ``authorization key'' must be 
specified. This key has to be distributed among the involved \program instances and scan definition files. 

For more details on this mode, see sec.~\ref{sec:gridcomputing}.

\paragraph{Test}
This mode can be used to test the formulas and other general directives given in the scan definition file. For this,
\program will process points in parameter space that are supplied directly by the user and only indirectly consider
parameter ranges specified in the scan definition file. 
Specifying points can either be done via the command line argument ``\cliarg{pars}'' (see sec.\ 
\ref{sec:CLIarguments}) or directly via a prompt on the terminal. Input via the terminal will be saved for 
later re-use and can be accessed via arrow up and down keys. Calculation results will also be saved. 
Additionally, the user can choose to run the calculation on one of the configured \program worker instances.

\subsection{General Concepts}\label{sec:concepts}


\subsubsection{Running of External Code}

To make it possible to run more than one process in parallel, \program will create temporary directories, in 
which each process will run. This means that, in general, external code must make \textbf{no}
assumptions on the location of the current directory and must be able to run from \textbf{any} directory on the 
computer. In addition, any side-effects outside of the current directory should be minimal -- this also means that 
external code should generally be what is usually referred to as ``re-entrant''.

\subsubsection{Structure and Storage of Results}
Every result of processing a parameter point consists of three parts:
\begin{description}
    \item[Parameters] A parameter is a value taken from a specified range. The parameters directly and uniquely 
        label every point in the parameter space.
    \item[Variables] A variable is a simple function of the parameters. They are calculated before substituting
        anything into the \textit{template file}. This can be used to re-parametrize the calculation,
        e.g.\ from $a, b$ to $a/b, a \cdot b$, or to obtain values following more complicated distributions than
        what is currently supported for parameters in \program.
    \item[Data] This denotes all values as derived and returned by the \textit{point processor}, see 
        sec.~\ref{sec:pointprocessor}.
\end{description}
Each line in the result file is then a tabulator separated list of values consisting of either the concatenation of 
these three sub-lists of values (customizable in ``scan'' mode) for valid points or only parameter values and 
exclusion reason for invalid ones.
Each point has to fulfill the following two criteria to be considered valid:
\begin{description}
    \item[No Error] The \textit{point processor} must have successfully returned, i.e.\ no error must have occurred
        (``Python has not raised an exception''). This usually also means that external programs must have exited 
        with an error code of 0.
    \item[Inside Bounds] The three sets of values of above have to satisfy a list of user-supplied 
        bounds/constraints.
\end{description}
The tabulator separated format can be read directly by gnuplot \cite{Gnuplot_4.4} and by Wolfram 
Mathematica\R \cite{Mathematica_7.0} (using its \code{Import} function).

As result from performing a scan defined in the file ``\textbf{$name$.scan}'', \program generates the
following files in the current working directory\footnote{One can change this with the command line argument 
\fncliarg{output\_dir}.}: 
\begin{description}
    \setlength{\itemsep}{0pt}
    \item[$name$.scan.log] Log file containing the most important messages from \program. (Mode: all)
    \item[$name$.scan.data] Valid result points. One line per point in parameter space,
        tabulator separated values. In ``optimize'' and ``explorer'' mode, one additional column is appended 
        containing the fitness function or likelihood value. (Modes: all except test, worker and mcmc)
    \item[$name$.scan.excluded-data] Excluded result points. Same format as ``.data'' files except for a single 
        column prepended containing only the character `E' for error/exclusion. The columns following that
        contain the parameter values and the reason for exclusion or error message as text. Note that upon 
        importing data formatted like this, \program will silently ignore the first column (`E').
        (Modes: all except test, worker and mcmc)
        
    \vspace{6pt}
    
    \item[$name$.scan.resume] Current resume position. (Mode: scan)
    \item[$name$.scan.speed] Current statistical information on the speed of the calculation. This will be 
        regenerated over time if deleted. (Mode: scan, mcmc)
        
    \vspace{6pt}
    
    \item[$name$.scan.testhistory] History of user supplied input points 
        in ``test'' mode. History file format of \code{readline} library. (Mode: test)
    \item[$name$.scan.testdata] Valid points obtained during calculations in ``test'' mode.
        Same format as ``.data'' file. 
        Likelihood value is appended to each row if defined.
        (Mode: test)
    
    \vspace{6pt}
    
    \item[$name$.scan.chain.$i$] Valid points obtained from the $i$'th Markov chain (starting from $i=0$). Lines 
        include likelihood and stay count at the end for every point in addition to parameters, variables and data. 
        Otherwise, same format as ``.data'' file. (Mode: mcmc)
    \item[$name$.scan.rejected.$i$] All valid but rejected data points that the $i$'th Markov chain (starting from 
        $i=0$) encountered, i.e. those that were discarded only due to random chance. Rows include the 
        likelihood at the end for every point in addition to parameters, variables and data. Otherwise, same 
        format as ``.data'' file. (Mode: mcmc)

    \vspace{6pt}
    
    \item[$name$.scan.work] Set of points that were determined to be calculated next in a previous run of 
        \program. This is used to resume calculations without the need to regenerate this information. 
        (Mode: explorer)
    \item[$name$.scan.boundary] Cache for the determination of boundary points, i.e.\ points that are not 
        invalid and have missing neighbor points. (Mode: explorer)
    \item[$name$.scan.projectedpoints] Parameter values (TSV format) of the set of points that were
        determined to have maximal $z$ value in each of the projections. This can be used to do the minimal amount
        of calculation necessary to re-calculate all projection graphs. This is only updated during state one 
        iterations. (Mode: explorer)
    \item[$name$.scan.projection.$i$] The $x$-$y$-$z_{\max}$ graph of the $i$'th projection (starting from $i=0$) 
        in TSV format (only updated in state one). This is also used to assess the difference of the 
        projection plains between successive state one iterations. (Mode: explorer)
    
    \vspace{6pt}
    
    \item[$name$.scan.population] Last used population in the differential evolution algorithm. The 
        points are contained as lines consisting of likelihood and parameter values (in that order) 
        separated by tabulator characters. (Mode: optimize)
    \item[$name$.scan.optimum] Parameter point that maximizes the likelihood function. Same format as
        ``.population'' files. (Mode: optimize)
    
\end{description}

\subsubsection{Formula Input}\label{sec:formulainput}
\program understands and uses formulas in the form of Python code in several places in the scan definition file. 
They are compiled to Python bytecode and evaluated according to the following rules:
\begin{itemize}
    \item Parameter, variable and data values of each point in parameter space can be accessed using 
        ``\code{pars.NAME}'', ``\code{vars.NAME}'' and ``\code{data.NAME}'', where \code{NAME} is a name as given 
        in the scan definition file. Alternatively, the syntax ``\code{list[$i$]}'', where \code{list} can be 
        \code{pars}, \code{vars} or \code{data}, can be used to access the value of the $i$'th value with $i$
        starting from 0. Note that it is not possible to have unnamed parameters or variables, while unnamed data 
        values are allowed but only at the end of the list of data values and are only accessible using 
        ``\code{data[$i$]}''. 
        
        In some cases, only a subset of these values can be available, for more details see the relevant 
        formula directive description. For rules on what constitutes a valid name, see the description of 
        \opt{par\_names} in sec.~\ref{sec:itemssetup}. 
        
    \item In Python formula expressions, you can use all built-in Python functions as well as several 
        functions and constants from the Python module ``math''\footnote{The full list of exposed \fncode{math} 
        functions is: sin, cos, tan, asin, acos, atan, atan2, exp, log, log10, sinh, cosh, tanh, asinh, acosh,
        atanh. The only exposed constant is \fncode{pi}. Other functions ``$f$'' from the \fncode{math} module can 
        be accessed as ``\fncode{math.}$f$''} simply by name. 
        
        One can provide access to more Python modules using the directive \opt{helper\_modules} in the \opt{setup} 
        section, see sec.~\ref{sec:setup}.
    
    \item The output of formulas is assumed to be a single value unless stated otherwise (being something else 
        leads to undefined behavior). If this is not the case, it will be clearly mentioned in the relevant scan 
        definition directive description.
        
    \item Formulas for variables are evaluated in the order in which they are saved to the result file. Variables 
        calculated first can not directly depend on variables coming later. However, to circumvent this limitation,
        \program provides the ``\code{remember}'' function to remember values computed in formulas before.
        
\vspace{\baselineskip}
\begin{lstlisting}[language=Python,emph={remember},title={Usage:}]
# setting values
remember(var=value)
# retrieving values
remember("var")
\end{lstlisting}
    \item All formulas are evaluated with Python's flag for ``future division'' enabled.
        This means that contrary to ordinary Python code the division of two integers using the operator `/' yields
        a floating point number and not an integer. Standard integer division can still be done using `//'.
        
        Example: ``\code{1 / 2}'' $\to 0.5$, \mbox{``\code{1 // 2}'' $\to 0$}.
        
    \item Exceptions/errors (e.g. division by zero or invalid indexing) occurring in formulas for variables and 
        bounds simply count as a reason for exclusion of the parameter point. Exceptions in all other formulas 
        lead to a complete stop of the running scan, unless stated otherwise. It is advised to make use of the 
        ``test'' mode to catch ill-defined formulas early before they generate large erroneously excluded data sets.
\end{itemize}

\vspace{\baselineskip}
\begin{lstlisting}[language=Python,emph={pars,vars,data,exp,sin},title={Examples for formulas:}]
# A_0 =
  pars.a0 * pars.m0
# likelihood for Higgs mass fit =
  exp(-0.5 * ((data.mh0 - 125.7) / 0.4) ** 2)
# P_ee (electron neutrino survival probability)
  1 - sin(2*pars.theta) ** 2 * sin(
    vars.deltam12sqr * pars.L / 4 / pars.E
  ) ** 2
# bound check for BR(mu -> e gamma)
  data.BR_muegamma < 2.4e-12
\end{lstlisting}

\subsubsection{Template File}\label{sec:templatefile}

The \textit{template file} is a file containing special text fragments that are replaced with (parameter or 
variable) values specific to the currently processed point in parameter space. The possible forms for these 
placeholder text fragments are similar to Unix shell variable interpolation and are given by:
\begin{description}
    \item[\code{\$valueidentifier}] This will be replaced with the value corresponding to the point in question. 
        \code{valueidentifier} can only contain letters, numbers, underscore and dot. The first character not 
        conforming to this terminates the placeholder specification.
        
        This provides access to parameters and variables of a point through the placeholder fragments 
        ``\code{\$pars.NAME}'' and ``\code{\$vars.NAME}'' with the same rules for \code{NAME} as in input 
        formulas. In addition, both sets of values together  can be accessed directly using ``\code{\$NAME}'', 
        where parameter names take precedence before variable names.
        
    \item[\code{\$\{valueidentifier\}}] Equivalent to the placeholder fragment ``\code{\$valueidentifier}'' 
        except that the identifier can also contain the square bracket characters `[' and `]'. This is 
        also useful if valid \code{valueidentifier} characters follow the placeholder fragment.
        
        Examples: \\
        ``\code{\$\{intpart\}.\$\{fracpart\}}'', ``\code{\$\{vars[0]\}}''
    
    \item[\code{\$\$}] This placeholder will be replaced with the letter `\$'.
\end{description}
In the following, \textit{substituted template file} will refer to the template file with all of its placeholders 
replaced with the values belonging to the current point in parameter space. 

\subsubsection{Point Processor}\label{sec:pointprocessor}

The \textit{point processor} is a Python module containing a function ``\code{main}'' with either one or three 
arguments, namely the path of a \textit{substituted template file} and parameter and variable values. 
This function can thus either have the signature 
\begin{lstlisting}[language=Python,emph={main}]
def main(template_file_path):
\end{lstlisting}
or (useful for \textit{processors} that handle parameters and variables directly; see 
e.g.~sec.~\ref{sec:Example_ChargedLeptons})
\begin{lstlisting}[language=Python,emph={main}]
def main(template_file_path, parameter_values, variable_values):
\end{lstlisting}
The ``\code{main}'' function must return an object of type ``list'' (or another iterable type) 
consisting of the derived values. 

For more advanced purposes, the \textit{point processor} module can additionally contain a function 
``\code{init}'', which is called with the following arguments: the folder of the scan definition file, the 
Python SafeConfigParser object containing the scan definition directives and the Python module containing all 
functions and classes of \program. It has the signature
\begin{lstlisting}[language=Python,emph={init}]
def init(definition_file_path, config_object, module):
\end{lstlisting}
This can be used to initialize any internal structures depending on specifics of the scan definition. For an 
example, see the source code of the ``\code{ExampleProcessor}'' \textit{point processor} provided with the \program 
package.

In practice, the \textit{point processor} module's \code{main} function should be as simple as possible and should 
directly call another process and then parse its output for the data relevant for the calculation.
Since each new process that is started during point processing, e.g.\ a shell or other language interpreter, 
increases the calculation time, this can lead to -- depending on the overall performance -- significant delays for 
large parameter spaces. It is thus recommended to try to minimize the amount of added complexity in the processing 
of a point.

For an overview of \textit{point processors} already included in \program, see sec.\ \ref{sec:providedprocessors}.

\section{How to define a Scan}\label{sec:scandefinition}

This section covers the scan definition file format of \program with all currently available directives.

\subsection{Definition File Format}\label{sec:definitionformat}

The scan definition file format is based on the one accepted and processed by the Python module 
SafeConfigParser \cite{PyConfigParser}, which is very similar to one of standard ``.ini'' files as used by 
Microsoft Windows\R.
Each file consists of one or more sections labeled by the line ``\opt{[section\_name]}''. Each section consists 
of lines of name, value pairs of the form ``\opt{name}=\code{value}'' or ``\opt{name}:\code{value}'', with names 
specific to that section. It is \textbf{not} necessary to put quotation marks around a string \code{value}. 
Trailing whitespace in names and leading and trailing whitespace in values as well as trailing whitespace on every 
line is removed, and multi-line values can be given by indenting lines after the one with ``\opt{name}''.
Additionally, all \opt{names} are case-sensitive and later specifications with the same \opt{name} and 
\opt{section} simply overwrite earlier ones.

Comments are given by lines starting with ``\code{\#}'' or ``\code{;}''. Additionally, comments in 
otherwise non-empty lines can only be made using ``\code{;}'', and only if there is whitespace in front of it. 

The SafeConfigParser module also supports text substitution. This means values can contain references to other 
values in the same section (or optionally in a special section called ``\opt{DEFAULT}''). These references are then 
replaced with the referred to (final) value for the full evaluation of the directives.
An example for the scan definition file syntax is given by:
\begin{lstlisting}[
language=config,emph={version,information,dir,template,point_processor,one_hundred_percent},moreemph={[2]section}
] 
# comments look like this
[section]
version : 1
information = This config file has one quite long 
  but uninformative multi-line value
dir = IGNORED
template = %(dir)s/file
point_processor = %(dir)s/FullCalculation.py
dir = folder
one_hundred_percent = 100%%
\end{lstlisting}
This replaces the text fragment ``\code{\%(dir)s}'' with ``folder'' in the directives \opt{template} and
\opt{point\_processor}, such that the \opt{template} directive now has the value
\mbox{``folder/file''}. Note that the value ``IGNORED'' is ignored since it has been overwritten by 
setting \opt{dir} a second time. Note that, to write actual `\code{\%}' characters, one must instead write 
`\code{\%\%}'.

Please keep in mind that text substitution into formulas leads to direct evaluation of the substituted text 
fragment by Python. Thus the value string ``\code{\%(x)s ** 2}'' with \opt{x} being set to ``-1'' is substituted 
by SafeConfigParser to ``\code{-1 ** 2}''. Due to operator precedence, Python evaluates this expression to $-1$.
In contrast, the formula ``\code{pars.x ** 2}'' is correctly evaluated to $+1$. 

Beyond the default SafeConfigParser features, \program also supports the inclusion of other files using the syntax:
\begin{lstlisting}[language=config]
@include PathToFile
\end{lstlisting}
This causes the file accessible under the path ``\code{PathToFile}'' to be parsed directly as is at the 
position of this statement. In particular, multi-line values can be started in the parent file and be continued 
in the included file. If the given path is not absolute, the file is looked for in the following folders (in order):
\begin{itemize}
    \item the current folder, from which \program was run.
    \item the folder containing the main scan definition file, i.e.~the one that was not subject to 
        \code{@include} somewhere and was given last on the command line if applicable.
    \item the folder containing the \program executable file.
    \item the folder containing the real \program file if it was launched through a symbolic link.
\end{itemize}

Note that, as usual in Unix shells, paths starting with ``\code{\textasciitilde}'' or 
``\code{\textasciitilde username}'' have this initial part replaced with the respective user's home directory.

\subsection{The ``setup'' Section}\label{sec:setup}
This section contains the directives concerning the general setup of the calculation.

\begin{description}
    \item[\opt{version}] (integer) This specifies which \program version the scan definition is written for. 
        Defaults to $1$. 
        
        In future versions of \program, this will be used to enable backwards-incompatible features
        (the default value will not change).
    \item[\opt{mode}] (string) Can be one of ``scan'', ``mcmc``, ``optimize'', ``explorer'', ``test'' or 
        ``worker''. Specifies which scanning strategy should be employed. Can be overwritten using the 
        ``\cliarg{mode}'' command line argument, see sec.~\ref{sec:CLIarguments}.
    \item[\opt{concurrent\_processors}] (integer) Specifies the number of concurrently running processes that 
        \program should use on the \textbf{running} computer. This does not affect worker instances running 
        on other computers. For more details, see sec.~\ref{sec:gridcomputing}.
        
        This defaults to the number of processor cores of the running computer\footnote{Note that for processors 
        with hyper-threading capability or similar, this may give the number of logical cores instead of physical 
        ones.}.
        
    \item[\opt{template}] (path) The path to the \textit{template file}. If this directive is not present, \program 
        will not perform the actions outlined in sec.~\ref{sec:templatefile} and the template argument passed to the
        \textit{point processor's} \code{main} function will be `\code{None}'.
        
        This directive follows the same search rules for paths as the ``\code{@include}'' statement. 
        
    \item[\opt{point\_processor}] (path) The path to the source code file of the \textit{point processor} that is 
        used to calculate points. Note that it must be implemented in Python (or as a wrapper written in Python) 
        and must be importable without errors. If not given, \program will only calculate variables and no data 
        values.
        
        This directive follows the same search rules for paths as the ``\code{@include}'' statement.
        
    \item[\opt{helper\_modules}] (string) List (separated by `:') of module names or paths to Python module files
        that shall be made available to the evaluation of input formulas. If given as path to a file, a module can 
        be accessed by its´bare file name, i.e.~without folder and extension.
        
        Each module path is resolved using the same search rules for paths as the ``\code{@include}'' statement.
\end{description}

\subsection{The ``parameter\_space'' Section}\label{sec:itemssetup}

This section can contain the following directives:

\begin{description}
    \item[\opt{par\_names}] (string) List of names (separated by comma) given to the parameters. Whitespace
        around list elements is stripped. Names can contain letters, numbers and underscore, but must not start 
        with a number or underscore, must not be empty or a Python keyword and must not be in the list multiple 
        times.
        
    \item[\opt{par\_$name$}] (range) The value range for the parameter called $name$. A valid range is 
        given by ``\code{definition [with options]}, where \code{options} can be a comma separated list of 
        ``\code{name=value}'' pairs while \code{definition} can be one of the following
        \begin{itemize}
            \item a finite list of numeric values, e.g.\ ``\code{1, 2, 3, 4}''. Intermediate values can be replaced
                with an ellipsis ``...'' or ``..'', with at least two values preceding and one following the     
                ellipsis, e.g.\ ``\code{1, 1.2, ..., 2}''. Such an expression \mbox{``\code{a, b, ..., c}''} will
                be expanded to the list of numbers starting at $a$ and going to $c$ with step size $b-a$. Example:
                \begin{equation*} 
                    \text{``\code{1, 1.2, ..., 2}''} \to \{1, 1.2, 1.4, 1.6, 1.8, 2\}
                \end{equation*}
                The values $a$, $b$ and $c$ will always be part of the value range even if $c$ is not exactly 
                encountered using this expansion.
            \item an interval definition as in ``\code{interval(a, b)}''. This range type supports the options 
                \code{count} and \code{distribution}. The latter can be either ``linear'' or ``log'' specifying
                whether the parameter or its logarithm shall be uniformly distributed. If ``\code{count}'' is given,
                the range is automatically translated into a finite range with appropriate equidistant (linear or
                log) spacing and as many points. Example:
                \begin{align*}
                    &\text{``\code{interval(1, 2) with count=5}''} \\
                    \to &\{1, 1.25, 1.5, 1.75, 2\}
                \end{align*}
            \item the definition of range with a Gaussian distribution with mean \code{mu} and standard deviation 
                \code{sigma} written as ``\code{normalvariate(mu, sigma)}''. This range type also supports the
                option ``\code{count}'', which automatically translates the range into a discrete and smoothly 
                spaced set of values also following a normal distribution. Example:
                \begin{align*}
                    &\text{``\code{normalvariate(1, 2) with count=11}''} \\
                    \to &\{-1.77, -0.94, -0.35, 0.14, 0.58, 1.00, 1.42, \\
                        &\quad 1.86, 2.35, 2.94\}
                \end{align*}
        \end{itemize}
        In ``mcmc'', all parameter ranges can also have the option ``\code{mcmc\_stepsize}''. For details, see 
        sec.~\ref{sec:mcmc_options}.
    \item[\opt{var\_names}] (string) Comma-separated list of names for the variables. Same rules and behavior as 
        for ``\opt{par\_names}''.
    \item[\opt{var\_$name$}] (formula) Formula for calculation of the variable called $name$.
        Follows the formula input format detailed in sec.\ \ref{sec:formulainput}. Can only depend on parameters 
        and constants. Variable formulas are evaluated in the order in which they appear in \opt{var\_names}.
        Errors (e.g.\ division by zero or invalid indexing) occurring during the evaluation of these formulas will
        lead to exclusion of the point.

\vspace{\baselineskip}
\begin{lstlisting}[language=config,emph={var\_M1,var\_Au33},title={Examples for variables:}]
# Bino mass M_1 = r_1 M_{1/2}
var_M1 = pars.r1 * pars.M12
# Trilinear coupling (A_u)_{33} = A_t y_t
var_Au33 = pars.A_t * (pars.m_t / 174)
\end{lstlisting}
    \item[\opt{data\_names}] (string) Comma-separated list of names for the data values. Same rules as 
        \opt{par\_names} and \opt{var\_names}. If the \textit{point processor} returns more values 
        than names are given here, the excess values can be accessed by their index $i$ (starting from 0) using 
        the syntax ``\code{data[$i$]}''.
    \item[\opt{bound\_count}] (integer) The number of user-defined bounds or validity checks performed for each 
        point in parameter space. Defaults to $0$.
    \item[\opt{bound\_$i$}] (formula) Formula for the $i$'th constraint check (starting from $i=0$) performed for 
        each (otherwise valid) point in parameter space. Should compute a value that can be interpreted as True or 
        False. Note that you must define at least as many bounds as specified in \opt{bound\_count}, while surplus
        bounds are ignored.
        
        The check is performed only \textbf{after} the point in parameter space has been handed over to the 
        \textit{point processor} and the calculation has finished, i.e.\ ``\code{data}'' values are always 
        available. The constraint checks are not evaluated if the \textit{point processor} itself has generated an 
        error.
        
        This has the same semantics regarding exceptions/errors as \opt{var\_$name$}, i.e.\ errors lead to 
        exclusion of the point.
\end{description}

\subsection{Scan Mode Specific Directives}

In addition to the generally applicable directives above, the ``scan'' mode has the following modified semantics 
and directives. Directives written as ``\optsec{section} \opt{name}'' refer to ``\opt{name}'' in the section 
``\code{section}''.
\begin{description}
    \item[\optsec{setup} \opt{unit\_length}] (integer) The number of points that are processed in one batch. This 
        means that for each scan iteration a set of points with length given by this directive of not yet calculated
        points is selected and handed over to all running \textit{point processor} processes.
        The Python multiprocessing module ensures that the set is evenly distributed across all of them. After 
        all points in the set have been processed, the results are saved to the ``.data'' or ``.excluded-data''
        files (depending on the validity of the specific point).
        
        It is generally advised to set this value high enough so that the waiting time for all processes to 
        finish is not large compared to the full processing time of the set of points in parameter space.
        
        Defaults to $100 \,\cdot\, $\code{concurrent\_processors}.
    \item[\optsec{parameter\_space} \opt{mode}] (string) Can be either ``grid'' (default), ``scatter'' or ``file''. 

        In ``grid'' mode, the usual interpretation of parameter definitions as described in 
        sec.\ \ref{sec:itemssetup} applies and all parameter ranges must be finite.
        
        In ``scatter'' mode, a random sample of points is drawn from the parameter ranges to be processed.
        The number of points can be adjusted using the directive \opt{point\_count} in the \opt{parameter\_space}
        section.
        
        In ``file'' mode, the parameter space is not given by the Cartesian product of parameter ranges, but by 
        pre-determined parameter space points given in the files in the directive \opt{files} of the 
        \opt{parameter\_space} section. In this case, the directives ``\opt{par\_$name$}'' in 
        the \opt{parameter\_space} section are interpreted as input formulas (as detailed in 
        sec.\ \ref{sec:formulainput}) that, however, only have access to the ``\code{file}'' values instead of
        ``\code{pars}'', ``\code{vars}'' or ``\code{data}''. These ``\code{file}'' values correspond to the 
        columns (tabulator separated) of each line in the files specified in the \opt{files} directive. They can 
        be accessed by index or via the names specified in the \opt{file\_columns} directive.
    \item[\optsec{parameter\_space} \opt{point\_count}] (integer) The number of points that should be calculated. 
        Only applies in parameter space mode ``scatter''.
    \item[\optsec{parameter\_space} \opt{files}] (string) List of file names separated by `:' (whitespace is 
        significant except for leading and trailing one of full lines!) to be used as definition of the parameter 
        space. Only applicable in ``file'' parameter space mode (or ``explorer'' mode). 
        
        If the command line parameter \cliarg{output\_dir} is set and a path is just a file name, it is first 
        looked for therein. Otherwise, each path is resolved using the same search rules as the 
        ``\code{@include}'' statement.
        Lines starting with a column containing only `E', i.e.\ lines containing excluded points, are ignored.
    \item[\optsec{parameter\_space} \opt{file\_columns}] (string) List of names (separated by comma) given to the 
        columns in the files given in the directive \opt{files}. Same rules and behavior as 
        ``\opt{par\_names}'', ``\opt{var\_names}'' and ``\opt{data\_names}''. Only applicable in ``file'' parameter 
        space mode.
    \item[\optsec{algorithm} \opt{out\_columns}] (formula) List of formulas separated by commas\footnote{
        Commas \textbf{inside} the formulas are handled correctly as long as all brackets and quotation marks are 
        properly closed and opened.} specifying what values should be saved to the ``.data'' file. Can be used for 
        e.g.\ re-ordering or re-parametrizing of the result values. 
        The formulas have access to ``\code{pars}'', ``\code{vars}'' and ``\code{data}'' (as well as ``\code{file}''
        values in ``file'' mode). 
         
        If not given, \program saves all data as usual (excluding ``\code{file}'' values).
\end{description}

\subsection{MCMC Mode Specific Directives}\label{sec:mcmc_options}

In addition to the general directives, this mode also has the following changed semantics in the sections 
\opt{setup} and \opt{parameter\_space} and its own directives in the \opt{algorithm} section. Directives written as 
``\optsec{section} \opt{name}'' refer to ``\opt{name}'' in the section ``\code{section}''.
\begin{description}
    \item[\optsec{setup} \opt{unit\_length}] (integer) The number of distinct points, i.e.\ not weighted by stay 
        count, each Markov chain shall find.
        
    \item[\optsec{setup} \opt{concurrent\_processors}] (integer) \program runs this many chains in parallel on the 
        local computer.
        
    \item[\optsec{parameter\_space} \opt{par\_$name$}] (range) In ``mcmc'' mode, all ranges can also have the 
        option ``\code{mcmc\_stepsize}'' that specifies the step size for the Gaussian proposal density around the
        parameter value of the last iteration. For discrete parameter ranges, this is understood to work on the
        indices into the list of values, i.e.\ the result of the proposal density sampling is rounded to the 
        nearest integer).
        
    \item[\optsec{algorithm} \opt{likelihood}] (formula) Formula for the calculation of the propagation 
        likelihood. This formula has to evaluate to non-negative numbers for all valid points in parameter space. 
        Its value is also saved together with the stay count in the resulting ``.chain.$i$'' files. Points with 
        vanishing \opt{likelihood} value are always rejected. To fit a \code{data} value to a measurement with mean 
        $\mu$ and uncertainty $\sigma$, the \code{likelihood} should be given by:
\begin{lstlisting}[language=config,emph={likelihood},mathescape=true]
likelihood = exp(
    -(data.x - $\mu$) ** 2 / (2 * $\sigma$ ** 2)
  )
\end{lstlisting}
\end{description}
Beyond fixing the length of all chains, \program does not perform any convergence analysis.

Note that this mode does not support distributing the calculation to more than one computer using the manager/worker
architecture. However, since Markov chains work autonomously, this can be done by hand by starting \program on each 
computer in ``mcmc'' mode separately.

\subsection{Optimize Mode Specific Directives}\label{sec:optimize_options}

This mode also has slightly adjusted semantics for the scan definition directives. Additionally, there are new 
directives in the \opt{algorithm} section, which make it possible to adjust the different parameters of the 
differential evolution algorithm \cite{StornPrice97}. Directives written as ``\optsec{section} \opt{name}'' refer 
to ``\opt{name}'' in the section ``\code{section}''.
\begin{description}
    \item[\optsec{setup} \opt{unit\_length}] (integer) The used population size. Defaults to $10 \, \cdot \, D$, 
        where 
        $D$ is the number of parameters with at least two possible values.
    \item[\optsec{algorithm} \opt{likelihood}] (formula) The fitness function that is maximized. No restrictions 
        are placed on its value beyond being a real number.
    \item[\optsec{algorithm} \opt{waiting\_threshold}] (float) The maximal absolute change in the fitness function 
        that is not considered convergent behavior. Defaults to $0$.
    \item[\optsec{algorithm} \opt{waiting\_threshold\_relative}] (float) The maximal relative change in the fitness 
        function that is not considered convergent behavior. Defaults to $10^{-\eta / 2}$, where $\eta$ is given 
        by the machine precision (16 in the case of double precision).
    \item[\optsec{algorithm} \opt{waiting\_time}] (integer) The number of iterations, during which the change in 
        the likelihood function value is below the one governed by the waiting thresholds, that the algorithm 
        should wait to make sure convergence has really taken place. Defaults to $10 \, \cdot \, D$, where 
        $D$ is the number of parameters with at least two possible values.
    \item[\optsec{algorithm} \opt{differential\_weight}] (float) The differential weight as used in differential 
        evolution, see sec.~\ref{sec:optimize_algorithm}. Defaults to $0.6$.
    \item[\optsec{algorithm} \opt{crossover\_probability}] (float) The cross-over probability as is used in 
        differential evolution, see sec.~\ref{sec:optimize_algorithm}. Defaults to $0.5$.
\end{description}
In summary, the fitness $f$ is considered to have converged if 
\begin{equation*}
    |f_i - f_{i+1}| < \epsilon + \rho |f_i|\;,
\end{equation*}
for at least \code{waiting\_time}$ + 1$ consecutive iterations $i$, where $\epsilon$ is determined by the 
directive \opt{waiting\_threshold} and $\rho$ by the directive \opt{waiting\_threshold\_relative}.
Note that, due to the random nature of the algorithm, this does not make a statement on the precision 
of the obtained optimal point or fitness.

\subsection{Explorer Mode Specific Directives}\label{sec:explorer_options}
The ``explorer'' mode has the following semantics and custom directives. Directives written as 
``\optsec{section} \opt{name}'' refer to ``\opt{name}'' in the section ``\code{section}''.
\begin{description}
    \item[\optsec{setup} \opt{unit\_length}] (integer) Generate this many random points as initial point set.
        During running, neighbors of (at most) this many points are calculated when no projections are done in 
        that iteration, i.e. state three. This directive has no consequence in projection-enabled iterations 
        (state one or two).
        
        Defaults to $10 \,\cdot\, D$, where $D$ is the number of parameters with at least two possible values.
        
    \item[\optsec{algorithm} \opt{loading\_filter}] (formula) Formula analogous to ``\opt{bound\_$i$}'' in the 
        section ``\code{parameter\_space}'' that can be used to specify a condition that points have to satisfy to 
        be loaded from disk into memory. Note that this only applies to loading, and points calculated afterwards 
        are not filtered by this. If not given, \program loads all points.
        
        Note that this may cause \program to calculate points twice.
        
    \item[\optsec{parameter\_space} \opt{files}] (string) List of file names separated by `:' (whitespace is 
        significant!) to be used as additional source for known parameter space points. \program will not write to 
        these files.
        
        If the command line parameter \cliarg{output\_dir} is set and a path given here is just a file name, it is 
        first looked for therein. Otherwise, each path is resolved using the same search rules as the
        ``\code{@include}'' statement. 
    \item[\optsec{algorithm} \opt{likelihood}] (formula) Formula for the calculation of the likelihood quality 
        criterion for all points. This can be negative.
        
    \item[\optsec{algorithm} \opt{min\_likelihood}] (float) The minimal likelihood value that a point in parameter 
        space has to have to be considered for further calculations. Points with values lower than this will still 
        be saved in the ``.data'' file, but will be treated as invalid for exploration.
        
    \item[\optsec{algorithm} \opt{likelihood\_steps}] (string) List of numerical values separated by comma
        specifying bins into which the likelihood of all points should be categorized. Example for bin interval 
        determinations:
        \begin{align*}
            &\text{``\code{1, 2.71, 3.141}''} \\
            \to &(-\infty, 1), [1, 2.71), [2.71, 3.141), [3.141, \infty)
        \end{align*}
        
        During an iteration in state three, \program only considers points in the highest likelihood, non-empty
        bin for further exploration. Once more than the three most likely bins are fully depleted of boundary 
        points to calculate neighbors for, points in the most likely bin are unloaded from memory - they still
        remain on disk in the ``.data'' file, but are `forgotten' to save memory\footnote{If you absolutely cannot
        tolerate duplicates in the result data set, you should probably not use this directive.}.
        If not given, the only likelihood bin is $(-\infty, \infty)$.
        
        This is only used in state three iterations.
    
    \item[\optsec{algorithm} \opt{disabled\_states}] (string) List of algorithm states (separated by comma, as
        integers) that shall not be used for the scan. Note that one cannot disable state three without 
        specifying projections.
        
    \item[\optsec{algorithm} \opt{projection\_count}] (integer) The number of projections defined in the
        scan definition file. Defaults to $0$. Note that you must define at least as many projections as 
        specified here, while surplus projections are ignored.
        
    \item[\optsec{algorithm} \opt{projection\_$i$}] (formula) Formula for the combined specification of $x$ 
        and $y$ value (separated by comma) to which points are projected. All projections are counted starting
        from $i=0$.
        
    \item[\shortstack{\optsec{algorithm} \opt{projection\_$i$\_x}\\
        \optsec{algorithm} \opt{projection\_$i$\_y}}] (formula) Separate formulas for $x$ and $y$ projection 
        coordinates. Either these two or \opt{projection\_$i$} can be used.
        
    \item[\optsec{algorithm} \opt{projection\_$i$\_z}] (formula) Formula for $z$ coordinate analogous to $x$
        and $y$. If this directive is not specified, it defaults to the likelihood as given in 
        \opt{likelihood} in the \opt{algorithm} section.
        
    \item[\optsec{algorithm} \opt{projection\_$i$\_filter}] (formula) Formula for a condition that points have to 
        fulfill to be considered for the calculation of the $i$'th projection. If not given, all valid points are 
        considered for the projection calculation.
        
    \item[\optsec{algorithm} \opt{extrapolated\_projections}] (string) The list of projection indices 
        (starting from 0 and separated by comma) of the projections for which interpolation and extrapolation 
        should be done to smoothen the $x$-$y$-$z_{\max}$ graph. If not given, all projections are interpolated 
        and extrapolated.
        
    \item[\optsec{algorithm} \opt{symmetry\_count}] (integer) The number of symmetry transformation applied to 
        projected points in state one iterations. Defaults to $0$. Note that you must define at least as many 
        symmetries as specified, while surplus symmetries are ignored.
        
    \item[\optsec{algorithm} \opt{symmetry\_$i$}] (formula) The list of transformation (sub-)rules (separated by 
        comma) of the form ``\code{par\_ident:formula}'', where ``\code{par\_ident}'' can be one of the
        following: an index (starting from 0) into the list of parameters, a (Python) string specifying the 
        parameter name, a Python identifier corresponding to the parameter name.
        The expression ``\code{formula}'' then gives the value the selected parameter should be changed to. 

\vspace{\baselineskip}
\begin{lstlisting}[language=config,emph={symmetry_0,symmetry_1,symmetry_2},title={Examples for symmetries:}]
# All of these symmetries do the same thing
#   (assuming par_names = x, y, phi)
symmetry_0= 2: pars.phi - pi, 0: -pars.x
symmetry_1= "phi": pars.phi - pi, "x": -pars.x
symmetry_2= phi: pars.phi - pi, x: -pars.x
\end{lstlisting}
        This transformation will then be applied to the points having maximal likelihood for (at least) one $x$-$y$ 
        point in one of the projection plains (in state one). The resulting points will then be calculated as well. 
        Note that resulting points that do not fall onto the grid and those where the symmetry transformations 
        generate an error, are ignored.

\end{description}

\section{Command Line Interface}\label{sec:CLIarguments}

\program can be started from the command line using the following structure and options:
{\ttfamily
\begin{description}
    \item[\code{\programfilename}] [-h] [-v] 
~[{-}-mode MODE] \\
~[-P] [{-}-pars VALUE [VALUE ...]] \\
~[-o output\_dir] [-p PORT] [input\_file]
\end{description}
}
\noindent Positional arguments:
\begin{description}
    \item[\code{input\_file}] Scan definition file specifying the parameter scan. If not given, \program will 
        show a menu with files to choose from. If multiple files are specified, they are read in sequence and 
        interpreted as one scan definition. The last file given determines the name applicable to output files
        such as ``.data'' files and others.
\end{description}
\noindent Optional arguments:
\begin{description}
    \item[\code{-h}, \cliarg{help}] Print out help message for the command line interface and exit.
    \item[\code{-v}, \cliarg{version}] Print out \program version and exit.
    \item[\cliarg{mode}] (string) Can be used to temporarily override the setting \opt{mode} of the \opt{setup}
        section in the scan definition file without changing it. This specifies which algorithm should be used.
        Possible values are ``scan'', ``mcmc'', ``optimize'', ``explorer'', ``worker'' or ``test''.
        
    \item[\code{-o}, \cliarg{output\_dir}] Directory in which data and other files for the requested scan should be
        written to. \program will create it if necessary. Defaults to the current working directory.
        
    \item[\code{-p}, \cliarg{port}] (integer) Can be used to temporarily override the setting given in 
        \opt{port} (in the \opt{setup} section) in the scan definition file without changing the file. 
        This specifies the port that \program uses to listen for requests in ``worker'' mode.
        
    \item[\cliarg{pars}] (string) Space-separated list of parameter values to be used in ``test'' mode
        (ignored in other modes). Multiple parameter points can be specified using ``\cliarg{pars}'' multiple 
        times. These points will be processed before any user input, if given.
        
    \item[\code{-P}, \cliarg{profiling}] If present, this causes \program to print out so called profiling 
        information in test mode if processing is done on the local computer. This only concerns code run in 
        Python\footnote{External code will only be listed as one big call to a function in the 
        ``\fncode{subprocess}'' Python module and will have no substructure.}, so it is only useful for more 
        sophisticated \textit{point processors}.
        
    \item[\code{-D}, \cliarg{debug}] Enable debug mode. This causes \program to output some additional information.
        
    \item[\cliarg{randomseed}] Use a specific seed for the pseudo-random number generator. This can be used to
        make calculations involving random numbers deterministic between different instances of \program.
\end{description}

Regardless of how the scan definition file is given, \program will first show an overview of the processed 
configuration before it will evaluate or calculate anything. If everything is in order and the user agrees, 
\program starts or resumes the scan. In ``scan'' and ``mcmc'' mode, the program will show an overview of the 
current overall progress, an estimated time of completion of the scan and some more possibly useful information.
In ``optimize'' and ``explorer'' mode, there is only a subset of such information available and completion 
estimation is only done where feasible, i.e.\ for sub-tasks with a defined length of calculation.

The user can interrupt any calculation at any time using the Ctrl+c key combination on the keyboard.
Calculations can then be resumed by launching \program again with the same \cliarg{output\_dir} and scan definition
file name. On resuming, \program will try to continue in exactly the state it was in before. In some cases, this 
may mean that points have to be calculated more than once, but \program will try its best that they will only be 
saved once in the result data set.

\textbf{Warning:} \program will not check whether the scan definition file changed while it was interrupted. 
Therefore, we advise caution to not accidentally mix up incompatible data sets.

\section{Using Multiple Computers}\label{sec:gridcomputing}

Every instance of \program that is running with a scan definition that has a non-empty \opt{workers} directive 
(in the \opt{setup} section) and a mode that is not ``worker'' is considered a manager instance. Manager instances 
delegate their calculation to the worker instances reachable by the addresses given in \opt{workers} in addition 
to the local processes governed by the value given in \opt{concurrent\_processors} in the \opt{setup} section. 
All workers are assigned work proportional to their respective \opt{concurrent\_processors} values.
The mode under which the manager instance runs is not relevant to worker instances as they only calculate data for 
points in parameter space requested by manager instances. Worker instances generally ignore specified parameter
ranges and process whatever parameter values they are given.

Worker instances can be used by multiple manager instances at the same time. However, they will return their 
results in the order in which they were requested, which can lead to significant delays when used by multiple 
manager instances. While running in ``worker'' mode, \program will show a list of the currently requested point 
batches and the last ten complete ones.

Note that worker instances do not notice when manager instances are interrupted and will continue calculating 
what they are tasked with, if not interrupted themselves.

The scan definition options relevant for the manager/worker architecture are the following (in the \opt{setup} 
section):
\begin{description}
    \item[\opt{workers}] (string) Comma-separated list of IP addresses or network names and ports under which 
        worker instances are reachable. Each list entry is expected to have the form \mbox{``address[:port]''}, 
        where the port defaults to 31415. \\
        Example:
        \begin{equation*}
            \text{\code{machineA, machineB:15707, 192.168.0.123}}
        \end{equation*}
        
    \item[\opt{port}] (integer) Specifies the port under which \program should listen in ``worker'' mode. Defaults 
        to 31415 if not present. The port must be between 1 and 65535 and must not already be in use.
        
    \item[\opt{authkey}] (string) Shared secret or authorization key shared among all instances (both manager and 
        workers) participating in the same calculation. This should be used to make sure scan definitions of 
        manager and worker instances are compatible with each other and that no unauthorized access is possible.
        
    \item[\opt{concurrent\_processors}] (integer) This number is specific to the running \program process 
        whether it is running in manager or worker mode and has no influence on other machines.
    
    \item[\opt{unit\_length}] (integer) If the default value is used for this directive and it depends on the 
        number of concurrent processes, it will be updated once a survey of all available worker instances has
        been done, to include the full number of processes.

\end{description}

\section{Included Point Processors}\label{sec:providedprocessors}

\subsection{The Minimal Processor ``ExampleProcessor''}

Very simple \textit{point processor}, which only demonstrates the structure of a typical \textit{point processor} 
and a few possibly useful functions and tricks.
Other than that, it has no configuration, does no calculation and returns an empty list of data values.

\subsection{The Simple Processor ``SimpleProcessor''}

This \textit{point processor} runs a user-supplied program with the \textit{substituted template file} as command
line argument and extracts all integer and floating point numbers that are not part of a word from the command's 
output\footnote{The number matching behavior can be tested from the shell using the command line: \newline
\mbox{\fncode{\textbf{\$} program | python SimpleProcessor.py}} \newline
which will echo the output of \code{program} and will mark matched numbers with indices from the front and 
back attached.}.
It is considered an error, i.e.\ leads to exclusion of a point, if the called program returns a non-zero 
exit code or takes too long. For a usage example, see sec.~\ref{sec:scalardm}.

This processor supports the following directives in its own \opt{SimpleProcessor} section:
\begin{description}
    \item[\opt{program}] (command line) Single command that shall be launched with the \textit{substituted 
        template file} as last argument in the directory containing it. If the first part of the given value is 
        just a name, it is first resolved using the \code{PATH} environment variable (so just like in a regular 
        shell). If it is not found this way, it is looked up using the same search rules as the ``\code{@include}'' 
        statement.
        
    \item[\opt{timeout}] (integer) Number of seconds the \textit{point processor} waits for the program to 
        complete its calculation. Defaults to $10$. Note that this applies \textbf{per point}.
        
    \item[\opt{data\_values}] (formula) Formula that evaluates to the data of interest contained in the list 
        of numbers extracted from the program's output.
        Can be a single or multiple values contained in a Python list-like object.
        This expression has access to the Python variables
        \code{pars} and \code{vars}, as well as the list \code{values}, which contains the extracted 
        numbers, and all mathematical functions and \opt{helper\_modules} as other formulas do.
        
\begin{lstlisting}[language=config,emph={data_values},title={Some examples:}]
data_values = values[0], values[10] / pars[0]
data_values = [values[i] for i in [1, 4, 9]] + 
              [2 * values[6]] 
\end{lstlisting}
\end{description}
The simple processor will show its derived configuration directly after its initialization.

\subsection{Analyzing SLHA Files with ``SLHAProcessor''}

This \textit{point processor} turns a \textit{substituted template file} over to a Susy Les Houches Accord 
(SLHA) \cite{Skands:2003cj} compliant spectrum generator or similar. It then reads and parses the resulting 
SLHA output files\footnote{The parser result can be reviewed using the command line: \newline
\mbox{\fncode{\textbf{\$} python SLHAProcessor.py file.slha}}
}, extracts a user-defined set of data values and returns them to \program.
For a usage example, see sec.~\ref{sec:higgsmass}.

It has the following custom scan definition directives in the \opt{SLHAProcessor} section:
\begin{description}
    \item[\opt{program}] (string) The command line used to process the template SLHA input file.
        The syntax corresponds to a basic subset of the Unix shell syntax. The supported features include full 
        quotation handling, input and output redirection using ``\code{< file}'' and ``\code{> file}'' and 
        chaining of commands using either `\code{;}', `\code{\&\&}' or new line characters as delimiter (however, 
        beware of misinterpretations of `\code{;}' as comments). In contrast to normal shell script, all three 
        delimiters are equivalent and execution is aborted as soon as a command returns a non-zero exit code. 
        
        The \textit{substituted template file} can either be referenced by path using the sub-string
        ``\code{\{template\}}'' as command line argument or will be passed to the standard output of the first
        command if referenced nowhere this way.
        
        All binaries are looked up following the same rules as the \opt{program} directive of the 
        \code{SimpleProcessor} and are executed in the directory of the \textit{substituted template file}.
        
\vspace{\baselineskip}
\begin{lstlisting}[language=config,emph={program},title={Examples for valid \opt{program} directives:}]
# running of &&-separated list like in shell
program = foo && bar && baz
# equivalent to the above:
program = foo; bar; baz
# multi-line also the same (and maybe cleaner)
program = foo
          bar
          baz

# input redirection is supported
# beware: ";" only begins comment if 
#  whitespace in front of it!
program = foo < infile > outfile; no comment
program = foo < infile > outfile ; comment

# redirections overwrite each other
program = foo > file1 > file2
# -> only file2 is created and written to

# file name patterns and variable expansion 
#  are not supported
program = echo *.scan $PATH > listing
# -> listing contains: "*.scan $PATH" 

# special characters and whitespace can be 
#  escaped just as in POSIX shells
program = three\ word\ program with_argument
program = "weird&&name" && html\<tags\>in_name

# if {template} is given as parameter, it is 
#  replaced with the path to the substituted 
#  template file upon execution
program = process_argument {template}

# if {template} is not used anywhere, it is
#  automatically fed to the stdin of the first
#  command
program = process_stdin
# is equivalent to
program = process_stdin < {template}
\end{lstlisting}

    \item[\opt{timeout}] (integer) Number of seconds the point processor waits for the program to complete its 
        calculation. Defaults to $10$. Note that this applies \textbf{per point}.
        
    \item[\opt{slha\_files}] (string) List of file names separated by `:' (whitespace is significant except 
        for leading and trailing one of full lines!) that shall be parsed as SLHA files for the selection and 
        evaluation of data.
        
    \item[\opt{slha\_data}] (formula) Python code that evaluates to the requested data contained in the SLHA 
        files specified in the \opt{slha\_files} directive. Can be a single or multiple values contained in a 
        Python list-like object.
        
        This formula expression has access to the same mathematical functions and constants as the formula 
        input detailed in sec.\ \ref{sec:formulainput} -- including the \code{pars} and \code{vars} variables 
        -- as well as to the object ``\code{slha}''.
        
        The ``\code{slha}'' object gives access to the parsed SLHA data in the following way:
\begin{lstlisting}[language=config,emph={slha,matrix}]
# access to BLOCKs with/without scale Q:
# slha[file_index][block, Q][index]
# slha[file_index][block][index]

# Higgs mass
slha[0]["MASS"][25]

# 3, 3 entry of up-type Yukawa matrix 
#  at or near scale Q=1000 GeV
slha[0]["YU", 1000][3, 3]
# "index-less" ALPHA block
slha[0]["ALPHA"][()]

# BLOCKs are also usable as Python matrices
# Careful: indices start from 0!
slha[0].matrix("YU")[i][j]
# (also includes an IMYU BLOCK if it exists)

# access to DECAY blocks
# slha[file_index]["DECAY"][pdgcode][index]

# gluino(1000021) decay width
slha[0]["DECAY"][1000021]["width"]
# gluino decay branching ratio to
#  2 particles, namely ~d_L(1000001) dbar(-1)
slha[0]["DECAY"][1000021][2, 1000001, -1]
\end{lstlisting}
        where \code{file\_index} is an index (starting from 0) into the list of parsed SLHA files specified in the 
        \opt{slha\_files} directive. 
        
        For BLOCKs, the index/value distinction is made such that (except for special cases\footnote{In the current 
        version of \program, these special BLOCKs are: FOBS, FOBSSM, FOBSERR, FMASS, FPARAM, FCONSTRATIO,
        HiggsBoundsInputHiggsCouplingsBosons and HiggsBoundsInputHiggsCouplingsFermions. For details, see 
        \cite{Bechtle:2013wla}, \cite{Mahmoudi:2010iz} and the source code of the SLHAProcessor module.}) 
        the last entry on each data line is treated as the value and the preceding entries as the 
        index. For DECAY blocks, the first entry in each data line is treated as the value and the rest as 
        index. If no index entries exist, e.g. in the case of the ``ALPHA'' block of the SLHA, the value can 
        be accessed via the index ``()'' (empty tuple).
        
        Block names are case insensitive and requesting a block with a specific scale Q finds 
        the block with the nearest value to the one requested, but gives a warning if it differs by more than 
        $1\%$ (only visible in ``test'' mode).
        
\end{description}
Thus, an example scan definition section that causes ``MySpectrumGenerator'' to be called with the
\textit{substituted template file} as its single command line argument, launches another analysis program
on the generated ``Spectrum'' file and then reads out some result values would look like the following:
\begin{lstlisting}[
    language=config,
    emph={program,input_mode,slha_files,slha_data},morekeywords={[1]SLHAProcessor}
]
[SLHAProcessor]
program = MySpectrumGenerator {template} > Spectrum 
    Analyser Spectrum > Analysis.SLHA
slha_files = Spectrum:Analyis.SLHA
slha_data = slha[0]["MASS"][25],
    slha[0]["MASS"][1000021],
    slha[0]["YU", 1000][3, 3],
    slha[0]["DECAY"][1000021]["width"],
    slha[0]["DECAY"][1000021][2, 1000001, -1],
    slha[1]["SOMECOMPLETELYDIFFERENTANALYSIS"][()]
\end{lstlisting}

The SLHA processor will show its derived configuration directly after its initialization.
Note that there is no check whether the number of names given in the \opt{data\_names} directive (in the 
\opt{parameter\_space} section) is consistent with the value of the \opt{slha\_data} directive.

\subsection{Multiple Point Processors with ``ProcessorChain''}

This processor can be used to combine two or more processors in the same scan. For this, it reads exactly
one directive in its own \opt{ProcessorChain} section.
\begin{description}
    \item[\opt{point\_processors}] (string) List of file names separated by `:' (whitespace is significant
        except for leading and trailing one of full lines!) of the \textit{point processor} modules that shall 
        be combined. The order is significant.
\end{description}

The \code{ProcessorChain} processor passes each point to the \textit{point processors} in the order in which 
they are given. The resulting list of data values is then the concatenation of the sub-data lists in that same 
order.

If one \textit{point processor} occurs multiple times in the list, both instances can be configured differently 
by prefixing the relevant scan definition section with ``ProcessorChain:$i$:'', where $i$ is the index of the 
relevant \textit{point processor} in the list (starting from 0). The \textit{point processor} will then see the 
prefixed section merged with all unprefixed sections.

\vspace{\baselineskip}
\begin{lstlisting}[
    title={Usage example:},
    language=config,
    emph={point_processors,name1,name2},
    morekeywords={[1]ProcessorChain,Processor}
]
[ProcessorChain]
point_processors = Processor:Processor

[Processor]
name1 = foo
name2 = bar

[ProcessorChain:0:Processor]
name1 = baz

[ProcessorChain:1:Processor]
name2 = baz

# -> first will see name1=baz, name2=bar
#    second will see name1=foo, name2=baz
\end{lstlisting}

\section{Example Scan Definitions}\label{sec:examples}

This section demonstrates how \program can be instructed to perform parameter scans for several types of 
calculations. It contains the following scans with their show-cased features:
\begin{description}
    \item[Scalar Dark Matter] ``test'' mode, ``scan'' mode for grids, usage of \code{SimpleProcessor}.
    \item[Lepton Mixing and Charged Lepton Corrections] ``mcmc'' and ``optimize'' modes, custom processors, 
        text substitution, ``\code{@include}'' statement, variables, helper modules.
    \item[Maximal Higgs Mass in mSUGRA] scattering scan mode, usage of \code{SLHAProcessor}, bounds, ``explorer'' 
        mode, parameter space mode ``file'', remote concurrency.
\end{description}
All files needed to perform the example scans are included in the \program package, except for publicly available 
external code -- for those, quick instructions for setting them up are included instead.
In addition, also the data files used for the figures and results shown are included within the \program package. 
For details on how to exactly reproduce them, see the included ``\textbf{README}'' file.

\subsection{Scalar Dark Matter}\label{sec:scalardm}
In this example, we reproduce the results of \cite{McDonald:1993ex} where the standard model (SM) is extended by a 
real scalar gauge-singlet $S$ that is supposed to be the single component of the dark matter (DM) relic density.
To this end, we make use of the $\mathbb{Z}_3$ symmetric dark matter model \cite{Belanger:2012vp} implemented in
\code{micrOMEGAs3.6.9.2} \cite{Belanger:2001fz,Belanger:2013oya} to calculate the dark matter relic density 
$\Omega_{DM}$ (``\code{Omega\_DM}'') and dark-matter-nucleon cross-sections $\sigma_{S,N}$ (``\code{sigma\_S\_N}'') 
for $N=p,n$ (proton and neutron) as a function of the scalar coupling $\lambda_{S1}$ (``\code{laS1}'') to the 
Higgs boson and the dark matter particle mass $M_S$ (``\code{MS}'').

Note that the model as implemented in \code{micrOMEGAs} actually includes a complex scalar gauge-singlet instead 
of a real one, so we have to divide $\Omega_{DM}$ by two to directly compare it with \cite{McDonald:1993ex}. 
To obtain the correct limit in all other aspects, the other scalar couplings of the $\mathbb{Z}_3$ model will be 
set to zero and the surplus particles will be decoupled by setting their masses to 1000 TeV. The Higgs boson mass 
will be fixed to 125.7 GeV.

For the calculation, we will use a slightly modified\footnote{Namely, the following options in the file ``main.c''
of the ``Z3MH'' folder were disabled: MASSES\_INFO, INDIRECT\_DETECTION, NEUTRINO, DECAYS, CLEAN. This leaves only
OMEGA and CDM\_NUCLEON enabled. A patch file containing this modification is included in the \program package.} 
version of the code included in the \code{micrOMEGAs} package. The relevant binary (``\code{main-Z3MH}'') expects a 
``\textbf{data.par}'' input file given as command line argument, which can be written as the template file 
``\textbf{data.par.template}'':
\begin{lstlisting}[ ]
Mh      125.7
Mdm1    $MS
Mdm2    1.0e6
MHC     1.0e6
muppS   0
la3     0
la2     0
laS     0
laS1    $laS1
laS2    0
laS21   0
sinDm   0
\end{lstlisting}
Since micrOMEGAs prints its results to the console, we choose to use ``SimpleProcessor''.
Looking at the usual output of the code, we see that $\Omega_{DM}$ is the $4$'th number printed and
the DM-nucleon cross-sections are given by the $8$'th and $4$'th number from the back (with a variable number of values
in between).

The scan definition file ``\textbf{ScalarDM.scan}'' is then given by:
\begin{lstlisting}[language=config,
    emph={par_laS1,par_MS,bound_0,bound_1,program,data_values},morekeywords={[1]SimpleProcessor}
]
[setup]
mode = scan
template = data.par.template
point_processor = processors/SimpleProcessor.py

[SimpleProcessor]
program = main-Z3MH
data_values = values[3], values[-8], values[-4]

[parameter_space]
par_names = laS1, MS
data_names = Omega_DM, sigma_S_p, sigma_S_n

par_laS1 = interval(0.01, 10) with count=200, distribution=log
par_MS = interval(10, 10000) with count=200, distribution=log
\end{lstlisting}
We will first test our scan definition by launching \program in ``test'' mode by running
\begin{lstlisting}[language=sh,escapechar=|,keywords={},emph={user@machine,dir$,$}]
$ |\programfilename|--mode test ScalarDM.scan 
\end{lstlisting}
As always \program will first show an overview of the settings as derived from the scan definition file before any 
calculation is run. It will then ask the user directly for parameter values of points that shall be calculated leading 
to console interactions like the following (user input in bold):
\begin{lstlisting}[language=HTML,escapechar=!]
![\dots]!
# Enter point (format: laS1, MS; as numbers or 'random'):
> !\textbf{0.5, 1100}!
# Parameters:
laS1 = 0.5, MS = 1100.0
# Calculation done after: 0.041191 s
# Data:
-------------------------------------------
Omega_DM :     0.111 | sigma_S_p: 1.779e-09 
sigma_S_n: 1.814e-09 | ---      :       ---
-------------------------------------------
# Output columns:
------------------------------------------------
1  :       0.5 | 2  :    1100.0 | 3  :     0.111 
4  : 1.779e-09 | 5  : 1.814e-09 | ---:       ---
------------------------------------------------
![\dots]!
\end{lstlisting}
If all our test input points yield satisfying results, we can launch \program in its configured mode using
\begin{lstlisting}[language=sh,escapechar=|,keywords={},emph={user@machine,dir$,$}]
$ |\programfilename|ScalarDM.scan
\end{lstlisting}
and after some cross-checks by \program with the user, the scan is performed without much interaction and saved to 
the file ``\textbf{ScalarDM.scan.data}'' (among others). 

\begin{figure}[t]
    \centering
    \includegraphics[width=\plotwidthSDM]{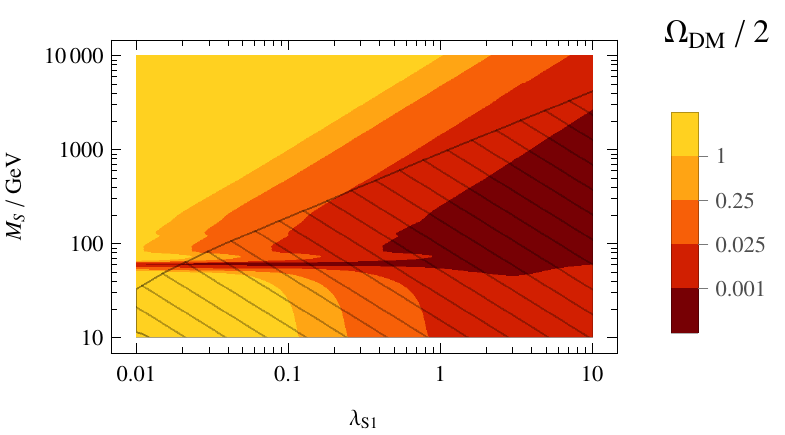}
    \caption{The DM relic density $\Omega_{DM}$ as function of the two scalar DM model parameters. The hatched
    region is excluded by $\sigma_{S,p}$ being above the 90\% CL bound as provided by the LUX experiment 
    \cite{Akerib:2013tjd} and obtained from DMTools \cite{DMTools}. Note that for better comparison
    with \cite{McDonald:1993ex}, the relic density has been divided by two.}\label{fig:scalarDMabundance}
\end{figure}

This file can then be read and turned into plots using the following code sketch for Mathematica
\begin{lstlisting}[language=Gnuplot,keywords={ListContourPlot,Import},emph=points]
points = Import["ScalarDM.scan.data", "Table"];
ListContourPlot[points[[All, {1,2,3}]]]
\end{lstlisting}
and for gnuplot
\begin{lstlisting}[language=Gnuplot]
splot "ScalarDM.scan.data" using 1:2:3
\end{lstlisting}
(both examples for $\Omega_{DM}$ over $\lambda_{S1}$-$M_S$ plots).
The resulting contour plot for the DM relic density is shown in fig.~\ref{fig:scalarDMabundance}. As we can see,
a comparison with fig.\ 2b in \cite{McDonald:1993ex} yields only very small (and expected due to e.g.\ a different 
Higgs boson mass) differences.

\subsection{Lepton Mixing and Charged Lepton Corrections}\label{sec:Example_ChargedLeptons}

In this example scan, we will demonstrate the ``optimize'' and ``mcmc'' scan modes. 
For this, we choose to investigate how well the present global fit of the PMNS matrix can be
reproduced by tribimaximal mixing in the neutrino sector and charged lepton mixing strictly 
between the first two generations. 

Since the computation for this is relatively easy, we write our own \textit{point processor} 
file ``\textbf{fit\_pmns.py}'':
\begin{lstlisting}[language=Python,moreemph={theta12e,delta12e,c12e,s12e,H,main,UTBM,Ue,
UPMNS,tan12,sin12,tan23,sin23,sin13,math,scipy,sp,dot,matrix,sqrt,cmath,cos,sin,sqrt,
exp,pars,template_file,vars}]
# import SciPy for nicer matrix usage
import scipy as sp
# import some more mathematical functions
from math import sin, cos, sqrt
from cmath import exp

def main(template_file, pars, vars):
    # we can also access pars and vars by name here
    theta12e = pars.theta12e
    s12e, c12e = sin(theta12e), cos(theta12e) 
    delta12e = pars.delta12e
    
    UTBM = sp.matrix([ 
        [-sqrt(2/3.), 1/sqrt(3),          0],
        [  1/sqrt(6), 1/sqrt(3),  1/sqrt(2)], 
        [  1/sqrt(6), 1/sqrt(3), -1/sqrt(2)] 
    ])
    Ue = sp.matrix([ 
        [c12e, s12e * exp(1j * delta12e),   0],
        [-s12e * exp(-1j * delta12e), c12e, 0],
        [0, 0, 1] 
    ])
    
    # m.H = hermitian transpose of m
    UPMNS = Ue.H * UTBM 
    
    # indices are 0-based!
    sin13 = abs(UPMNS[0,2])
    tan12 = abs(UPMNS[0,1] / UPMNS[0,0])
    tan23 = abs(UPMNS[1,2] / UPMNS[2,2])
    sin12 = tan12 / sqrt(1 + tan12 ** 2)
    sin23 = tan23 / sqrt(1 + tan23 ** 2)
    
    return [sin12, sin23, sin13]
\end{lstlisting}
As one can see, the code simply multiplies the tribimaximal mixing matrix \code{UTBM} with the 
charged lepton mixing matrix \code{Ue} and extracts the $\sin$'s of the three mixing angles.

The scan definition file for finding the best fit point (using the 2014 
NuFIT values \cite{GonzalezGarcia:2012sz} as target) is then given by (as file 
``\textbf{ChargedLeptons\_optimize.scan}''):
\begin{lstlisting}[language=config,moreemph={par_theta12e,par_delta12e,waiting_threshold,waiting_time,chi_squared}
]
[setup]
# do optimization with default population size 
#  (10*2 = 20 in this case)
mode = optimize

# use our custom processor 
#  (it uses no template file)
point_processor = fit_pmns.py

[parameter_space]
# two parameters that are angles in radians
# (only interested in first quadrant for theta)
par_names = theta12e, delta12e
par_theta12e = interval(0, 1.5707963267948966)
par_delta12e = interval(0, 3.141592653589793)

data_names = sin12, sin23, sin13

[algorithm]
# fit to "NuFIT Free Fluxes + RSBL"
#  (use higher minimum for theta23)
chi_squared = (
        ((data.sin12 ** 2 - 0.304) / 0.012) ** 2 +
        ((data.sin23 ** 2 - 0.577) / 0.03) ** 2 +
        ((data.sin13 ** 2 - 0.0219) / 0.0010) ** 2
    )
likelihood = -(%(chi_squared)s)
\end{lstlisting}
Note that, as introduced in sec.~\ref{sec:definitionformat}, the placeholder string ``\code{\%(chi\_squared)s}'' is 
replaced with the value of the directive \code{chi\_squared}.
Being interested in the probability distributions of our parameters, we also calculate those
using the Markov chain Monte Carlo algorithm (mode ``mcmc'') using the scan definition file 
``\textbf{ChargedLeptons\_mcmc.scan}'' (which extends the previous definition file using ``\code{@include}''):
\begin{lstlisting}[language=config,moreemph={par_theta12e,par_delta12e,chi_squared}]
@include ChargedLeptons_optimize.scan

[setup]
# do MCMC analysis with chains containing 10000 
#  points each
mode = mcmc
unit_length = 10000

[parameter_space]
# amend parameters with step sizes
par_theta12e = interval(0, 1.5707963267948966) with mcmc_stepsize=0.003
par_delta12e = interval(0, 3.141592653589793) with mcmc_stepsize=0.04

[algorithm]
# switch from a chi^2 to a distribution function
likelihood = exp(-0.5 * (%(chi_squared)s))
\end{lstlisting}
Here, we used the information
on the $\chi^2$ parabolas
obtained from the ``optimize'' scan to get an estimate for the MCMC 
step sizes.

\begin{table}
    \begin{center}\begin{tabular}{ccc}
        \toprule
         & best fit value & $1\sigma$ uncertainty \\
        \midrule
        $\theta_{12}^e$ in $^\circ$ & 12.07 & $^{+0.25}_{-0.31}$ \\
        $\delta_{12}^e$ in $^\circ$ & 74.7 & $\pm 5.1$ \\
        \midrule
        $\sin^2\theta_{12}^\text{PMNS}$ & 0.304 & $^{+0.12}_{-0.11}$\\
        $\sin^2\theta_{23}^\text{PMNS}$ & 0.4888 & $^{+0.0006}_{-0.0004}$\\
        $\sin^2\theta_{13}^\text{PMNS}$ & 0.0218 & $\pm 0.0010$\\
        \midrule
        $a$ in $10^{-5}$ & 1.36 & $^{+0.04}_{-0.03}$\\
        $b$ in $10^{-4}$ & 1.26 & $\pm 0.03$\\
        $c$ in $10^{-4}$ & 5.881 & $^{+0.008}_{-0.007}$\\
        \bottomrule
    \end{tabular}\end{center}
    \caption{Statistical information obtained from the fit and MCMC analysis of charged lepton corrections to 
    tribimaximal mixing. The Pearson correlation coefficient between the fitted parameters $\theta_{12}^e$ and 
    $\delta_{12}^e$ is $0.05$ and the minimal $\chi^2$ is $8.64$. The correlation coefficients between $a$, $b$ and 
    $c$ are $\rho_{ab} = -0.998$, $\rho_{ac} = 0.82$ and $\rho_{bc} = -0.80$.
    }\label{tab:ChargedLeptons}
\end{table}

In a final step, we will use the obtained information on $\theta_{12}^e$ and values for the electron and muon 
Yukawa couplings $y_e$ and $y_\mu$ \cite{Antusch:2013jca} to fit the parameters of a very simple flavor model.
For simplicity, we neglect mixing to the third generation and only consider the first two generations. The 
relevant Yukawa coupling matrix is then given by
\begin{equation}
    Y_e = \begin{pmatrix} 0 & b \\ a & c \end{pmatrix} \;,
\end{equation}
which is defined to appear in the Lagrangian as $L_i (Y_e)_{ij} e^c_j H_d$.
The relevant equations $Y_e$ has to fulfill to reproduce the Yukawa couplings and the mixing angle are 
given by
\begin{equation}
    |\det Y_e| = y_e \cdot y_\mu \;, \;
    || Y_e ||^2 = y_e^2 + y_\mu^2 \;, \;
    \left| \frac{b \cdot c}{a^2 + c^2} \right| = \tan \theta_{12}^e \;,
\end{equation}
where we used a zeroth order approximation in $y_e / y_\mu$ in the last equation and assumed all parameters to be 
real. Because the equation solving capabilities of standard Python are limited, we will again use the 
\code{SciPy} package. Since there is nothing else to calculate, we will omit the specification of a 
\textit{point processor} and instead do the whole calculation using variables.
The scan definition file ``\textbf{ChargedLeptons\_matrixfit.scan}'', which solves the relevant equations is then 
given by
\begin{lstlisting}[language=config,moreemph={par_ye,par_ymu,par_theta12e,var_a,var_b,var_c,Y_e,helper_modules}]
[setup]
helper_modules = scipy.optimize:scipy.linalg

[parameter_space]
par_names = ye, ymu, theta12e
# numbers taken from arXiv:1306.6879
par_ye = normalvariate(2.8501e-6, 0.0022e-6)
par_ymu = normalvariate(6.0167e-4, 0.0044e-4)
# ... and previous scans
par_theta12e = normalvariate(0.211, 0.005)

var_names = a, b, c
Y_e = [[0, b], [a, c]]
var_a = remember(a_b_c=scipy.optimize.fsolve(
          lambda (a, b, c): (
            abs(scipy.linalg.det(%(Y_e)s)) -
              (pars.ye * pars.ymu),
            scipy.linalg.norm(%(Y_e)s) ** 2 -
              (pars.ye ** 2 + pars.ymu ** 2),
            abs(b * c / (a ** 2 + c ** 2)) -
              tan(pars.theta12e)
          ),
          (2e-5, 1e-4 , 6e-4)
        ))[0]
var_b = remember("a_b_c")[1]
var_c = remember("a_b_c")[2]

mode = scatter
point_count = 100000

\end{lstlisting}
As can be seen, we used the \code{remember} function to cache the result of numerical equation solving to avoid
doing it again for each variable.

Finally, all resulting statistical information of all three parameter scans is summarized in 
tab.~\ref{tab:ChargedLeptons}. 

\subsection{Maximal Higgs Mass in mSUGRA}\label{sec:higgsmass}

In this example, we use the program \code{SoftSUSY3.5.1} \cite{Allanach:2001kg} to scan the mSUGRA parameter
space and find the maximal possible Higgs boson mass over certain subspaces. The results are then compared with 
the ones found in the study in \cite{Arbey:2011ab}.

We start by writing a \textit{template file} in SLHA format \cite{Skands:2003cj}. For the parameter scan, we 
intend to vary all 4+1 mSUGRA parameters and use the same input parameters as in \cite{Arbey:2011ab}. Thus we 
arrive at a file named ``\textbf{mSUGRA.slha.template}'' with the following content:
\begin{lstlisting}[language=config,moreemph={}]
Block MODSEL		    # Select model
    1   1                 # sugra
Block SMINPUTS		    # Standard Model inputs
    1	1.279160000e+02	  # alpha^(-1) SM MSbar(MZ)
    2   1.166370000e-05	  # G_Fermi
    3   1.184000000e-01	  # alpha_s(MZ) SM MSbar
    4   9.119000000e+01	  # MZ(pole)
    5	4.190000000e+00	  # mb(mb) SM MSbar
    6   1.729000000e+02	  # mtop(pole)
    7	1.777000000e+00	  # mtau(pole)
Block MINPAR		    # Input parameters
    1   $m0               # m0
    2   $m12              # m12
    3   $tanBeta          # tan beta at MZ, DRbar
    4   $sign_mu          # sign(mu)
    5   $A0  	          # A0
Block SOFTSUSY  # SOFTSUSY-specific 
    1   1.0e-03   # Numerical precision
    2   0.0	      # Quark mixing parameter
    5   1.0       # 2-loop soft mass/trilinear RGEs
\end{lstlisting}
Note that we disabled quark and general flavor mixing to make the analysis of superpartner masses not unnecessarily
complicated.

For the parameter scan, we write the following scan definition to a file with the name 
\mbox{``\textbf{HiggsMass\_scatterscan.scan}''}:
\begin{lstlisting}[language=config,moreemph={par_m0,par_m12,par_A0,par_sign_mu,par_tanBeta,bound_0,bound_1,bound_2,
program,slha_files,slha_data},morekeywords={[1]SLHAProcessor}]
[setup]
mode = scan

# use our template and the SLHAProcessor
template = mSUGRA.slha.template
point_processor = processors/SLHAProcessor.py

[SLHAProcessor]
# command line syntax as per SoftSUSY manual
program = softpoint.x leshouches < {template} > SLHASpectrum

slha_files = SLHASpectrum

# extract several masses from the spectrum file
#  (for the PDG code meanings, see data_names)
slha_data = slha[0]["MASS"][25],
 slha[0]["MASS"][1000022],slha[0]["MASS"][1000024],
 slha[0]["MASS"][1000006],slha[0]["MASS"][2000006], 
 slha[0]["MASS"][1000021], 
 slha[0]["MASS"][1000001],slha[0]["MASS"][1000002], 
 slha[0]["MASS"][2000001],slha[0]["MASS"][2000002],
 min([
   slha[0]["MASS"][code] 
   for code in [1000012, 1000014, 1000016]
 ]),
 slha[0]["MASS"][2000011],slha[0]["MASS"][2000013], 
 slha[0]["MASS"][1000015],slha[0]["MASS"][1000005]

[parameter_space]
# mSUGRA has 5 parameters:
par_names = m0, m12, A0, sign_mu, tanBeta

# scan over the following continuous ranges:
par_m0      = interval(50, 3000)
par_m12     = interval(50, 3000)
par_A0      = interval(-9000, 9000)
par_tanBeta = interval(1, 60)
# (except for the sign of mu of course)
par_sign_mu = -1, 1

# give names to all the masses 
data_names = mh0, 
    mN1, mC1, 
    mstop1, mstop2, 
    msG, 
    msdL, msuL, 
    msdR, msuR, 
    msNu, 
    mseR, msMuR, 
    msTau1, msb

# impose three constraints
bound_count = 3
# MSUSY must be below 3 TeV
bound_0 = sqrt(data.mstop1 * data.mstop2) < 3000
# masses have to satisfy current PDG bounds
bound_1 = data.mN1 > 46
    and (data.mC1 > 94 or pars.tanBeta > 40
      or data.mC1-data.mN1 < 3)
    and (data.msNu > 94 or pars.tanBeta > 40
      or data.mseR-data.mN1 < 10)
    and data.mseR > 107
    and (data.msMuR > 94 or pars.tanBeta > 40
      or data.msMuR-data.mN1 < 10)
    and (data.msTau1 > 81.9
      or data.msTau1-data.mN1 < 15)
    and min(data.msuL, data.msdL, data.msuR,
      data.msdR) > 1.11e3
    and (data.msb > 89 or data.msb-data.mN1 < 8)
    and (data.mstop1 > 95.7
      or data.mstop1 - data.mN1 < 10)
    and data.msG > 800
# the neutralino shall be the LSP
bound_2 = data.mN1 <= min(
    data[i] for i in range(len(data)) if i != 0
  )

# do a random scan
mode = scatter
point_count = 100000

\end{lstlisting}
As can be seen, we perform a ``scatter'' scan where we calculate $100000$ points taken at random from
a set of continuous or discrete ranges. For the bounds, we stuck to the constraint $M_{\text{SUSY}} < 3$ TeV as 
also imposed in \cite{Arbey:2011ab} and used the bounds as given in \cite{PDG2014} for the superpartner masses.

\begin{figure}[b]
    \begin{center}
    \includegraphics[width=\plotwidthHMOneD]{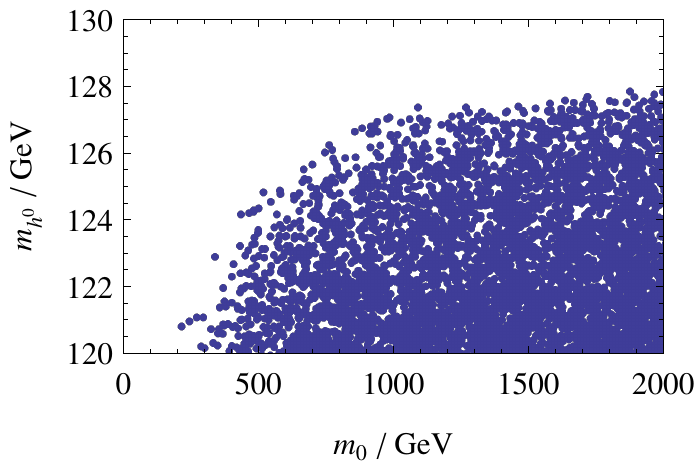}
    \includegraphics[width=\plotwidthHMOneD]{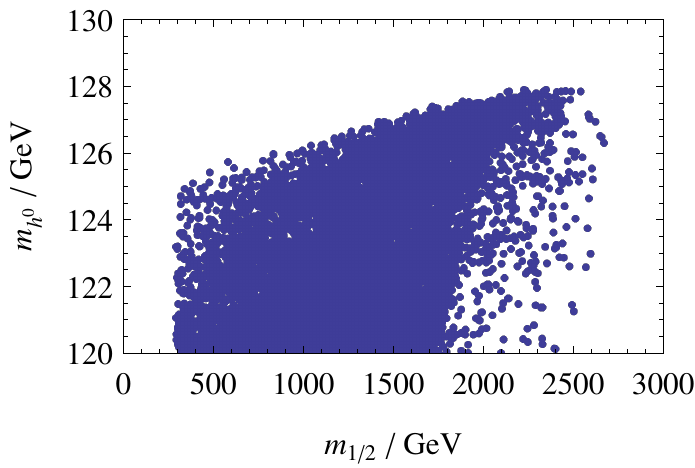}
    
    \includegraphics[width=\plotwidthHMOneD]{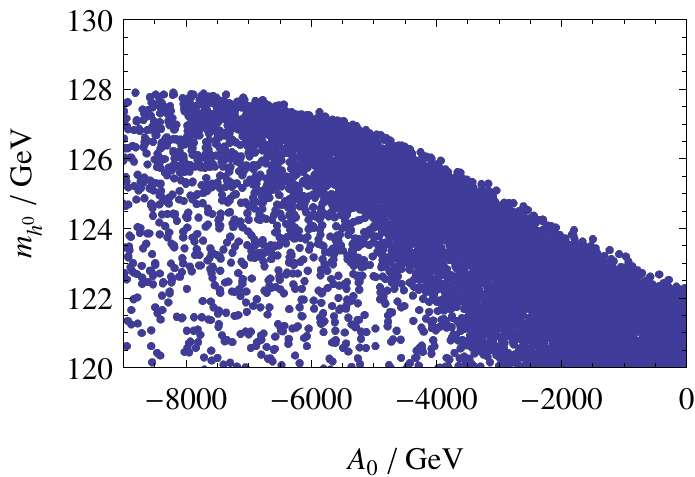}
    \includegraphics[width=\plotwidthHMOneD]{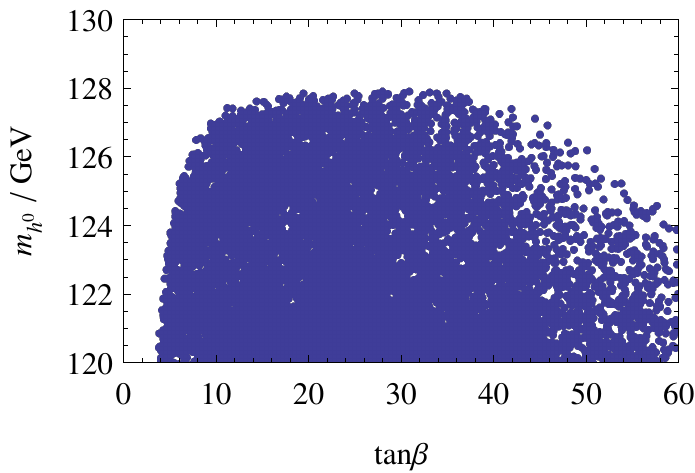}
    \end{center}
    \caption{Value of the Higgs boson mass $m_{h^0}$ as function of each of the four continuous mSUGRA parameters 
    (marginalizing over the others each time) as found in the scattering scan. No further constraints than the 
    ones in the scan definition file were applied.}\label{fig:HiggsMass_scatter}
\end{figure}

We then let \program run the parameter scan, which is complete after about 105 minutes (on a 3.4 GHz Intel i7 quad
core processor). 
The resulting plots can be seen in fig.~\ref{fig:HiggsMass_scatter}.
A comparison with the analogous plots of \cite{Arbey:2011ab} (fig.~3 a-d therein) shows that both have almost the
same behavior with only minor differences that can be attributed to more rigorous constraints for the superpartner 
spectrum in \cite{Arbey:2011ab}.

\begin{figure}
    \begin{center}
    \includegraphics[width=\plotwidthHMOneD]{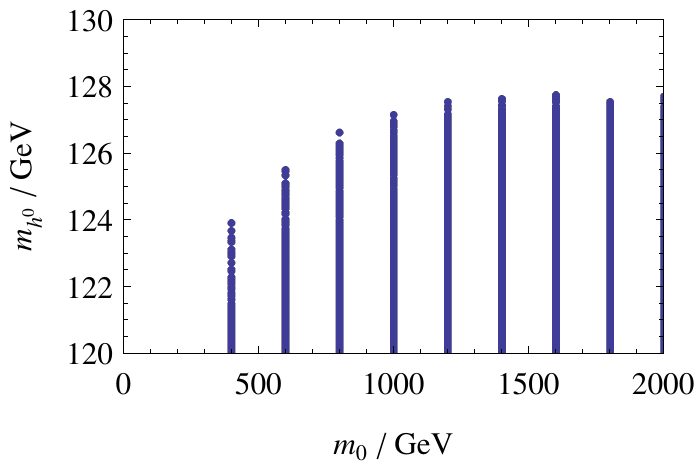}
    \includegraphics[width=\plotwidthHMOneD]{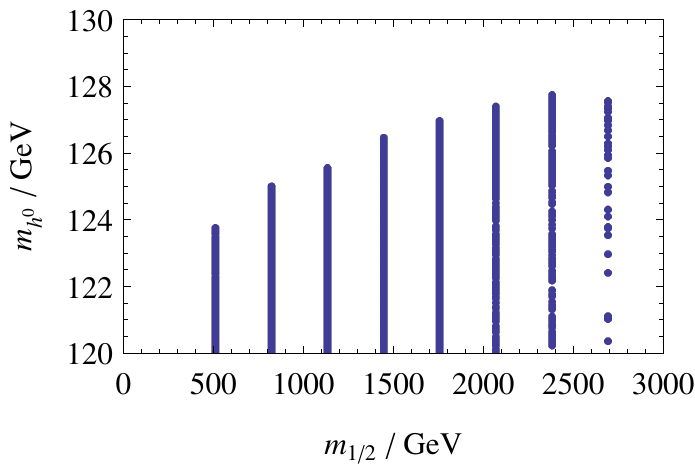}
    
    \includegraphics[width=\plotwidthHMOneD]{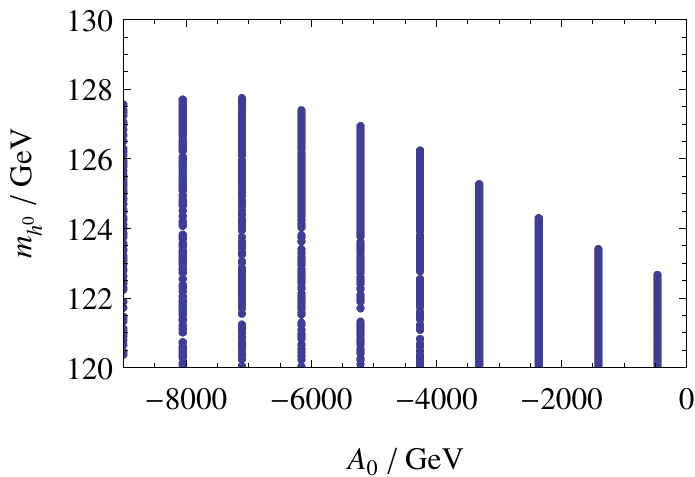}
    \includegraphics[width=\plotwidthHMOneD]{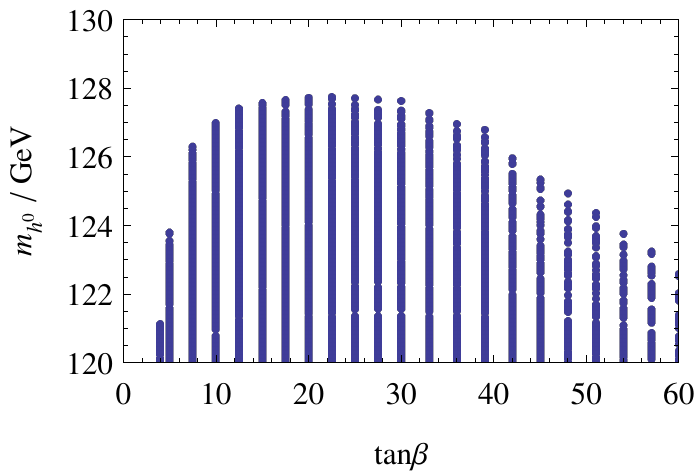}
    \end{center}
    \caption{Value of the Higgs boson mass $m_{h^0}$ as function of each of the four continuous mSUGRA parameters 
    (marginalizing over the others each time) as found in the grid scan. No further constraints than the ones in 
    the scan definition file were applied.}\label{fig:HiggsMass_grid}
\end{figure}

As an additional refinement, one might be interested in the maximal Higgs boson mass over two-dimensional parameter
subspaces instead of one-dimensional ones. This would involve binning the randomly distributed points
appropriately in the respective plain. Instead, we go for an automatically binned strategy involving a discrete
grid over which we scan our five mSUGRA parameters. The corresponding scan definition file 
``\textbf{HiggsMass\_gridscan.scan}'' only has a few lines:
\begin{lstlisting}[language=config,moreemph={par_m0,par_m12,par_A0,par_tanBeta}]
@include HiggsMass_scatterscan.scan
[parameter_space]
mode = grid

par_m0 = interval(200, 2000) with count = 10
par_m12 = interval(200, 3000) with count = 10
par_A0 = interval(-9000, 9000) with count = 20
par_tanBeta = 1, 2,..., 5, 7.5,..., 30, 33,..., 60

\end{lstlisting}
Again after about 95 minutes\footnote{Differences to the scattering scan are to be expected since the
parameter ranges slightly differ between scatter and grid scan.} (on the same 3.4 GHz Intel i7 quad
core processor), \program has finished the scan.
The resulting plots for one-dimensional subspaces are shown in fig.~\ref{fig:HiggsMass_grid} with (as expected) 
good agreement with the previously obtained data set.

\begin{figure}
    \begin{center}
    \includegraphics[width=\plotwidthHMOneD]{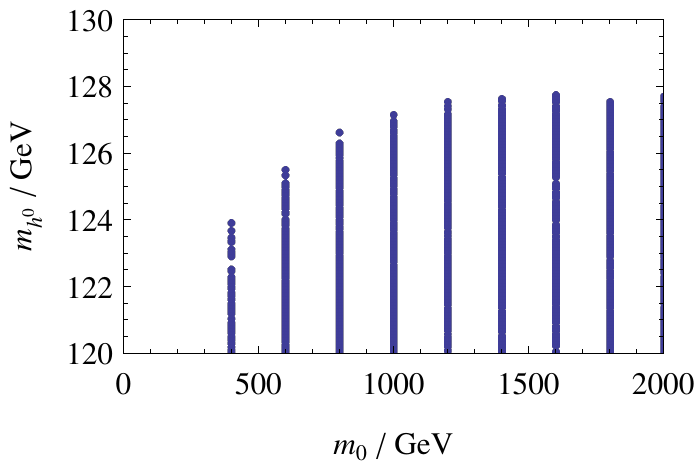}
    \includegraphics[width=\plotwidthHMOneD]{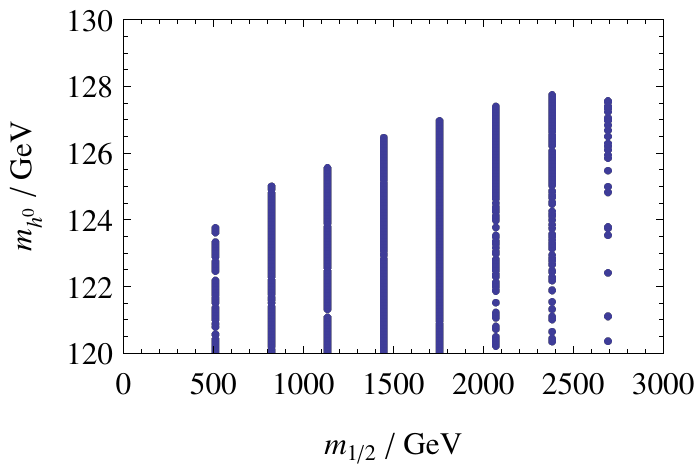}
    
    \includegraphics[width=\plotwidthHMOneD]{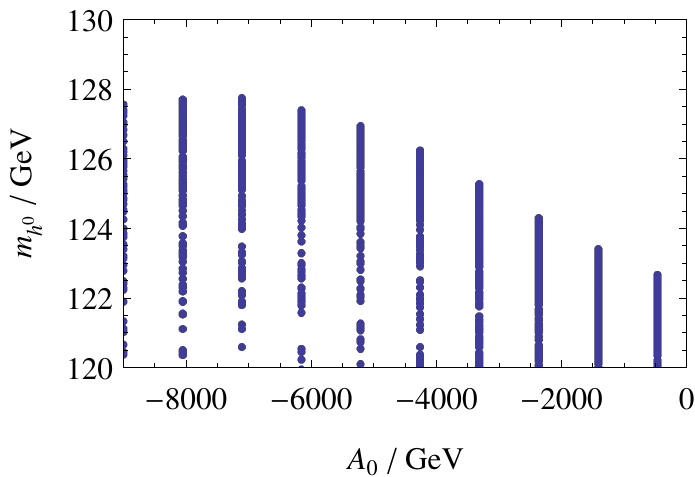}
    \includegraphics[width=\plotwidthHMOneD]{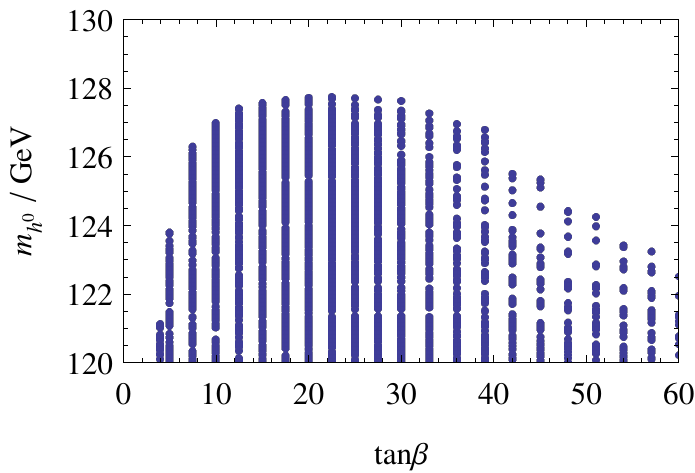}
    \end{center}
    \caption{Value of the Higgs boson mass $m_{h^0}$ as function of each of the four continuous mSUGRA parameters 
    (marginalizing over the others each time) as found in the ``explorer'' scan. No further constraints than the 
    ones in the scan definition file were applied.}\label{fig:HiggsMass_explorer}
\end{figure}

\begin{figure}
    \begin{center}
    \includegraphics[width=1.3\plotwidthHMTwoD]{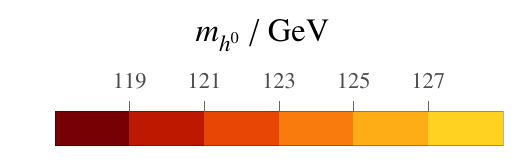}
    
        \includegraphics[width=\plotwidthHMTwoD]{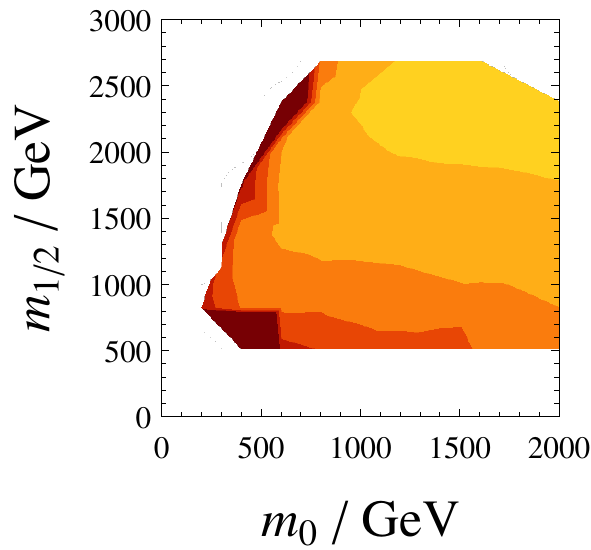}
    \includegraphics[width=\plotwidthHMTwoD]{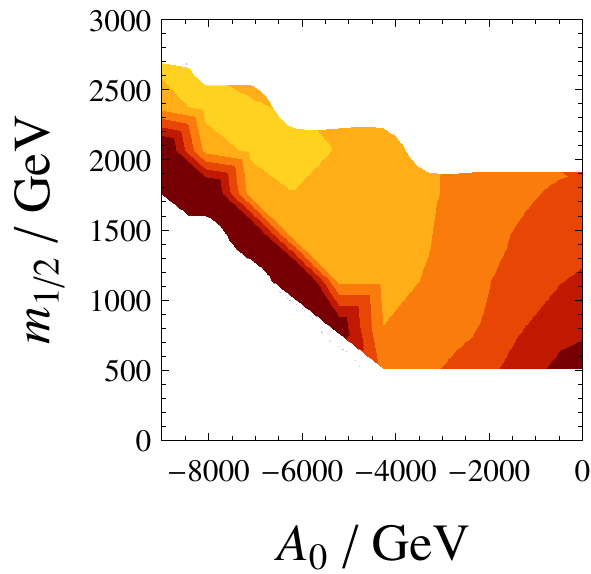}
    \includegraphics[width=\plotwidthHMTwoD]{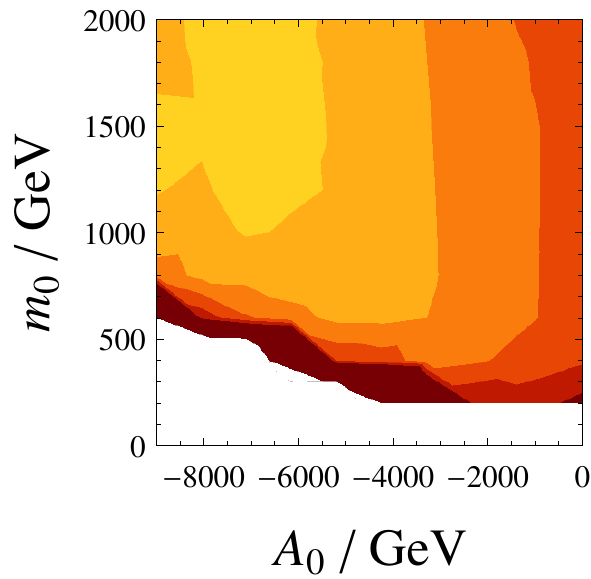}
    \includegraphics[width=\plotwidthHMTwoD]{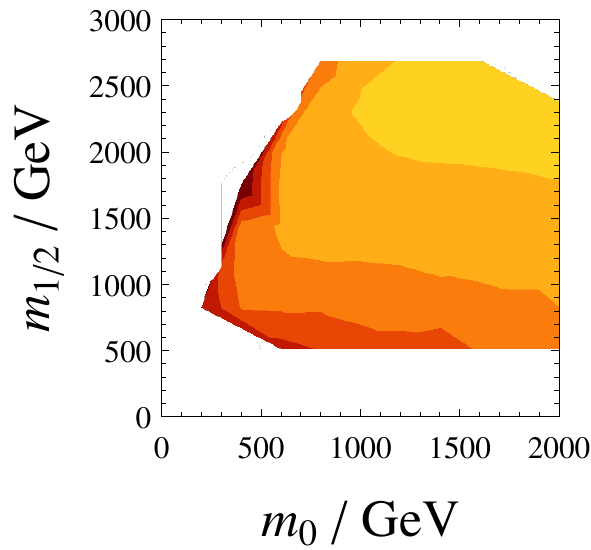}
    \includegraphics[width=\plotwidthHMTwoD]{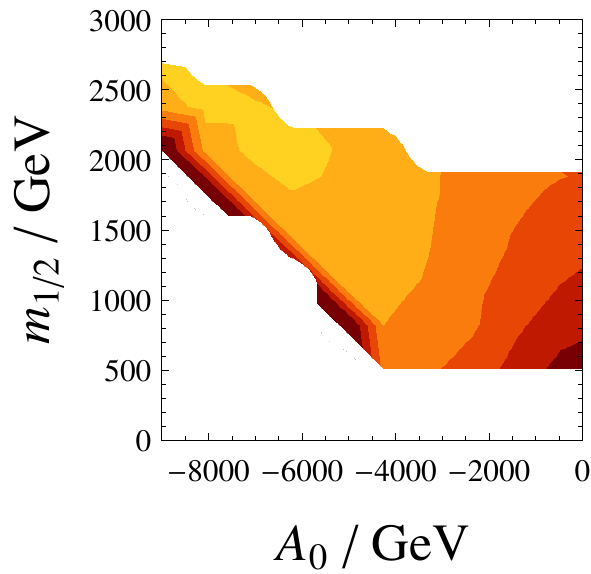}
    \includegraphics[width=\plotwidthHMTwoD]{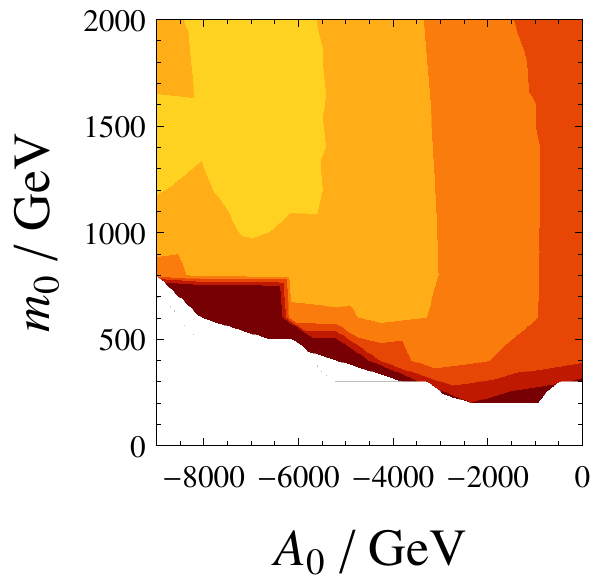}
    
    \end{center}
    \caption{Maximal value of the Higgs boson mass $m_{h^0}$ as function of the four continuous mSUGRA parameters 
    (marginalizing over the others) as found in the grid (upper) and ``explorer'' (lower) scan.
    }\label{fig:HiggsMass2}
\end{figure}

An additional feature of \program is the so called ``explorer'' mode. This mode is exactly designed for this
sort of calculation, i.e. maximizing a given function over several different plains in parameter space. Adapting
our grid scan to this mode is not complicated and yields the following scan definition file 
``\textbf{HiggsMass\_explorer.scan}'':
\begin{lstlisting}[language=config,moreemph={projection_0,projection_1,projection_2,projection_3,projection_4,
projection_5,projection_6,symmetry_0,symmetry_1,disabled_states}]
@include HiggsMass_gridscan.scan
[setup]
mode = explorer

[algorithm]
# maximize the Higgs boson mass!
likelihood = data.mh0
min_likelihood = 118

# calculate maximal Higgs boson mass over several 
#  different planes (or lines)
projection_count = 7
projection_0 = pars.m0, pars.m0
projection_1 = pars.m12, pars.m12
projection_2 = pars.A0, pars.A0
projection_3 = pars.tanBeta, pars.tanBeta

projection_4 = pars.m0, pars.m12
projection_5 = pars.m0, pars.A0
projection_6 = pars.m12, pars.A0

# two (probably not very good) symmetries
symmetry_count = 2
# no dependence on sign of mu?
symmetry_0 = sign_mu: -pars.sign_mu
# no dependence on sign of A0?
symmetry_1 = 2: -pars[2]

# only do state one calculation
disabled_states = 2, 3

\end{lstlisting}
After running this scan for about 5 minutes,
we find analogous plots to the two previous scans as shown in fig.~\ref{fig:HiggsMass_explorer}. Additionally, we
can compare plots of the maximal Higgs boson mass over some two-dimensional plains as shown in 
fig.~\ref{fig:HiggsMass2}. As can be seen, the differences are already very small compared to the calculation of 
the full parameter space grid.

Note that there are multiple ways to improve the data obtained using the ``explorer'' mode. For 
example, one can simply enable algorithm states two and three and let \program run longer. On the other hand, 
keeping in mind that in this example the starting points were chosen at random, one can also choose 
those more carefully. This can be done by hand using the ``test'' mode and consequently using the 
``.testdata'' file as input or by running a scan in the ``optimize'' mode beforehand and thereby creating an 
already nearly optimal starting point set.

As final part of this example, we will compare the results obtained with \code{SoftSUSY} with what we would 
have gotten with \code{SPheno3.3.3} \cite{Porod:2011nf,Porod:2003um}. For demonstration purposes, we will make
use of the manager-worker architecture, too. The content of the scan definition file 
``\textbf{HiggsMass\_recalc.scan}'' is then given by:
\begin{lstlisting}[language=config,moreemph={par_m0,par_m12,par_A0,par_sign_mu,par_tanBeta,program},
morekeywords={[1]SLHAProcessor}]
@include HiggsMass_explorer.scan
[setup]
mode = scan
workers = 127.0.0.1
authkey = HiggsMass_recalc

[SLHAProcessor]
program = SPheno {template} SLHASpectrum

[parameter_space]
mode = file
files = HiggsMass_explorer.scan.data
file_columns = m0, m12, A0, sign_mu, tanBeta, mh0
par_m0 = file.m0
par_m12 = file.m12
par_A0 = file.A0
par_sign_mu = file.sign_mu
par_tanBeta = file.tanBeta

[algorithm]
out_columns = pars.m0, pars.m12, pars.A0, 
    pars.sign_mu, pars.tanBeta, data.mh0 - file.mh0

\end{lstlisting}
Note that the used version of \code{SPheno} does not make use of exit codes to signal invalid parameters. 
However, it will write an incomplete SLHA spectrum file instead, which will lead to errors from 
\code{SLHAProcessor} due to a missing ``MASS'' block.

\begin{figure}
    \begin{center}
        \includegraphics[width=0.48\textwidth]{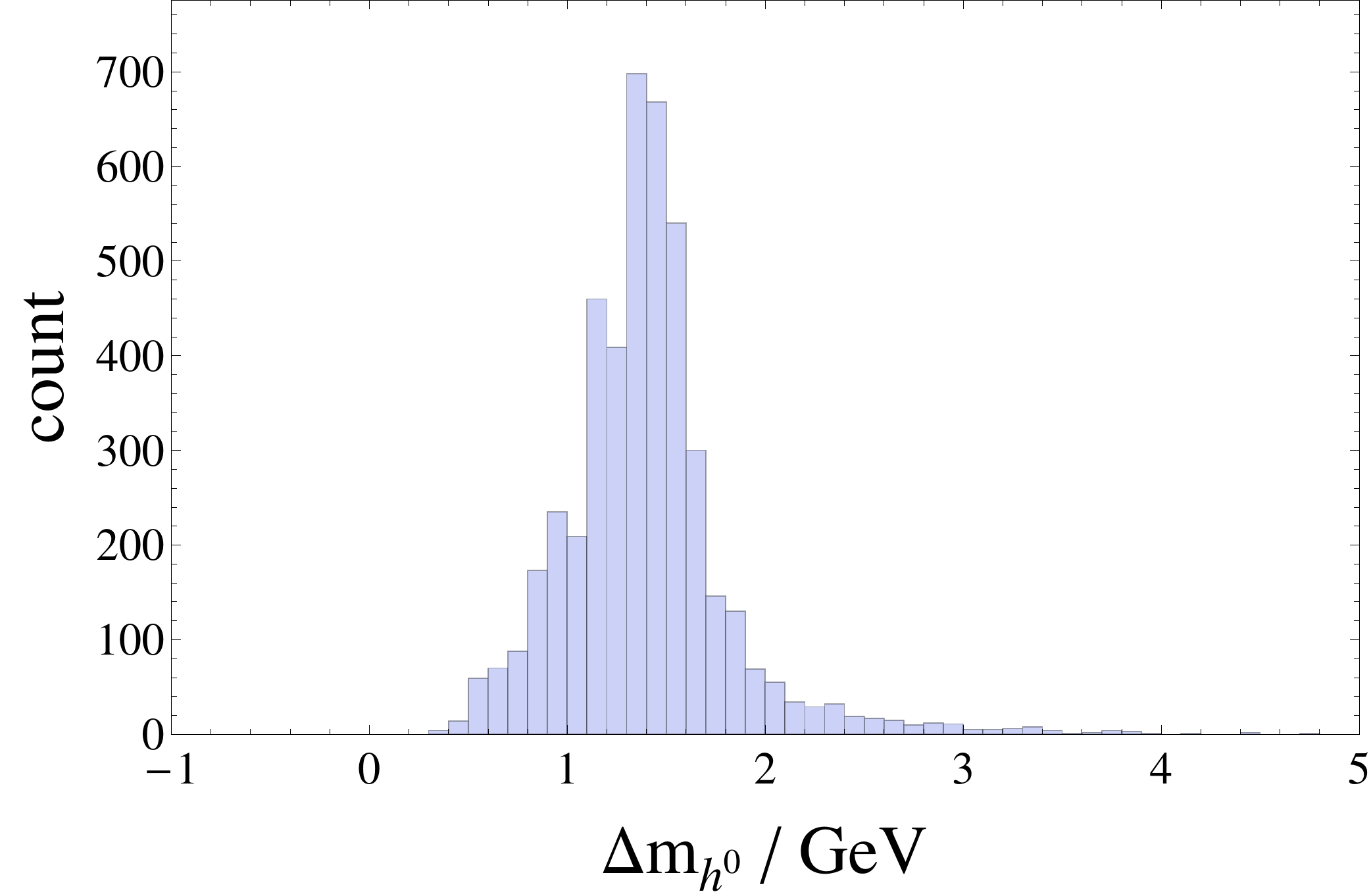}
    \end{center}
    \caption{Difference between the Higgs mass as calculated by the codes \code{SoftSUSY} and \code{SPheno} for 
    the points found in the ``explorer'' scan.
    }\label{fig:HiggsMassComparison}
\end{figure}

As can be seen, we use the parameter space mode ``file''. This means that the parameter space is taken 
directly from the points contained in the output file of the ``explorer'' scan. The parameter values are 
simply taken one-to-one from the data file. The \opt{out\_columns} directive causes that only the specified 
values are saved to the result file instead of all parameter and calculated data values. Finally, the worker 
instance of \program can be launched using
\begin{lstlisting}[language=sh,escapechar=|,keywords={},emph={user@machine,dir$,$}]
$ |\programfilename|HiggsMass_recalc.scan --mode worker
\end{lstlisting}
As shown in fig.~\ref{fig:HiggsMassComparison}, both codes only differ by about $1.5$ GeV in their Higgs mass 
result.

\section*{Acknowledgments}
This work is supported by the Swiss National Science Foundation. We thank Stefan Antusch, Ivo de Medeiros 
Varzielas, Oliver Fischer, Christian Gross and Constantin Sluka for useful discussions.

\newpage
\appendix
\renewcommand*{\thesection}{\Alph{section}}

\section{Implemented Algorithms}\label{app:algorithms}

\subsection{Markov Chain Monte Carlo}\label{sec:mcmc_algorithm}

The implemented Metropolis-Hastings algorithm \cite{MetropolisEtAl53,Hastings70} -- starting from a point \code{p0} 
-- can be sketched as:
\begin{lstlisting}[language=pseudocode,emph={}]
While count_points < requested_count_points:
    p1 = new_random_point_around(p0, stepsizes)
    If p1 is excluded by prior:
        stay_count = stay_count + 1
        Redo
    Else If p1 is excluded by constraints:
        stay_count = stay_count + 1
        Redo
    End
    L1 = L(p1) * prior(p1)
    L0 = L(p0) * prior(p0)
    
    a = random_uniform(0, 1)
    If a < min(L1 / L0 * q(p0, p1) / q(p1, p0), 1):
        Save p0 (with stay count)
        p0 = p1
        stay_count = 1
        count_points = count_points + 1
    Else:
        stay_count = stay_count + 1
    End
End
\end{lstlisting}
where \code{L(p)} is the propagation likelihood value for the point \code{p} and \code{q(p1, p0)} is the proposal 
density (up to normalization) for the point \code{p1} around \code{p0}. In the case of flat prior probability, 
\code{L} corresponds to the target probability distribution function (up to normalization).
For the relevant scan definition directives, see sec.~\ref{sec:mcmc_options}.

\subsection{Optimization using Differential Evolution}\label{sec:optimize_algorithm}

The implementation of the differential evolution algorithm \cite{StornPrice97} as applied to a fitness function 
\code{f} can be sketched as:
\begin{lstlisting}[language=pseudocode,emph={with,Find}]
While waiting $\le$ max_waiting_time:
    For each x in population:
        a, b, c = Find three random points 
            in population with 
            a $\ne$ b $\ne$ c $\ne$ a and a, b, c $\ne$ x
        xs = a + S * (b - c)
        For i from 1 to count_dimensions:
            r = random_uniform(0, 1)
            If r < rho:
                y[i] = x[i]
            Else:
                y[i] = xs[i]
            End
        End
        If f(y) > f(x):
            Replace x -> y in population
        End
    End
    
    new = max{f(x) for x in new population}
    old = max{f(x) for x in old population}
    If abs(new - old) < eps + eps_rel * old:
        waiting = waiting + 1
    Else:
        waiting = 0
    End
End
\end{lstlisting}
where \code{\normalsize S} $\in [0, 2]$ is the differential weight (default: 0.6), 
\code{rho} $\in [0, 1]$ is the crossover probability (default: 0.5), \code{eps} is the absolute waiting 
threshold (default: 0) and \code{eps\_rel} is the relative waiting threshold (default: $10^{-8}$ 
for double precision) -- the population size defaults to $10$ times
the number of parameters with at least two possible values. For more information on the choice of these algorithm 
parameters, see e.g.~\cite{Storn96}. For the relevant scan definition directives, see 
sec.~\ref{sec:optimize_options}.

Note that, after the last iteration, one has to extract the point with highest fitness value from the last 
population.

\subsection{Swarm-like Explorative Optimization}\label{sec:explorer_algorithm}

Each iteration of the algorithm starting from a known point set $S$ can be sketched as:
\begin{lstlisting}[language=pseudocode,emph={Process}]
If projections defined and state < 3:
   For each projection in projections:
      For each p in $S$:
         If state > 1 or p $\notin$ boundary($S$):
            Next
         End
         x, y, z = projected_coordinates(
            projection, p
         )
         If z > z_max(x, y):
            p_max(x, y) = p
            z_max(x, y) = z
         End
      End
      If state = 1:
         For each x, y in projected_plain:
            Add neighbors(best_point[x, y]) to $\mathcal{P}$
            For dx, dy in simple_orbit(0):
               Add points from line segment             
                  [p_max(x, y) p_max(x+dx, y+dy)] 
                  to $\mathcal{P}$ 
               Add points encountered by extending 
                  previous line segment to $\mathcal{P}$
               Apply symmetries to p_max(x, y) 
                  and add result to $\mathcal{P}$
            End
         End
      End
   End
Else:
   For each p in $S$:
      If $\exists$ n with n $\in$ neighbors(p) and n $\notin S$
         Add p to boundary bin $\mathcal{B}_i$ according to L(p)
      End
   End
   $\mathcal{B}$ = highest likelihood, non empty $\mathcal{B}_i$
   For each p in $\mathcal{B}$:
      Add neighbors(p) to $\mathcal{P}$
   End
End
If $\mathcal{P}$ is empty:
   If state < 3:
      state = state + 1
   End
Else:
   For each p in $\mathcal{P} \setminus S$:
      Process p and add it to $S$
   End
   If projections defined:
      state = 1
   End
End
$\mathcal{P}$ = {}
\end{lstlisting}
where \code{L(p)} is the likelihood function value for the point \code{p}. 
For the relevant scan definition directives, see sec.~\ref{sec:explorer_options}.

\bibliographystyle{elsarticle-num}
\bibliography{manual}{}
\end{document}